\documentclass[fleqn,usenatbib]{mnras}

\usepackage{subcaption}
\usepackage{float}
\usepackage{newtxtext,newtxmath}
\usepackage[normalem]{ulem}

\usepackage[T1]{fontenc}
\usepackage{multirow}

\usepackage{graphicx}	
\usepackage{amsmath}	
\usepackage{color}
\usepackage{natbib}
\usepackage{url}
\usepackage{placeins}

\definecolor{amethyst}{rgb}{0.6, 0.4, 0.8}

\definecolor{rr}{rgb}{1.0, 0.4, 0.4}

\definecolor{orange}{rgb}{1, 0.5, 0.}

\definecolor{verde}{rgb}{0., 0.6, 0.}
\definecolor{red}{rgb}{0.6, 0., 0.}
\newcommand{\rev}[1]{{#1}}

\newcommand{\revdue}[1]{{#1}}

\title[\rev{Dust evolution with MUPPI in cosmological volumes}]{\rev{Dust evolution with MUPPI in cosmological volumes}}

\author[Massimiliano Parente et al.\ ]{Massimiliano Parente $^{1,2}$\thanks{E-mail: mparente@sissa.it}, Cinthia Ragone-Figueroa$^{3,2}$, Gian Luigi Granato$^{2,3,4}$, Stefano Borgani$^{5,2,4,6}$, \newauthor Giuseppe Murante$^{2,4}$, Milena Valentini$^{8,9,2}$, Alessandro Bressan$^{1}$, Andrea Lapi$^{1,4,6,7}$
\\
$^{1}$ SISSA, Via Bonomea 265, I-34136 Trieste, Italy \\
$^{2}$ INAF, Osservatorio Astronomico di Trieste, via Tiepolo 11, I-34131, Trieste, Italy \\
$^{3}$ Instituto de Astronom\'ia Te\'orica y Experimental (IATE), Consejo Nacional de Investigaciones Cient\'ificas y T\'ecnicas de la\\ Rep\'ublica Argentina (CONICET), Universidad Nacional de C\'ordoba, Laprida 854, X5000BGR, C\'ordoba, Argentina\\
$^{4}$ IFPU - Institute for Fundamental Physics of the Universe, Via Beirut 2, 34014 Trieste, Italy\\
$^5$ Dipartimento di Fisica dell' Universit\`a di Trieste, Sezione di Astronomia, via Tiepolo 11, I-34131 Trieste, Italy
\\
$^6$ INFN - National Institute for Nuclear Physics, Via Valerio 2, I-34127 Trieste, Italy\\
$^7$ INAF - Istituto di Radio-Astronomia, Via Gobetti 101, I-40129 Bologna, Italy\\
$^{8}$ Universit{\"a}ts-Sternwarte, Fakult{\"a}t f{\"u}r Physik,  Ludwig-Maximilians Universit{\"a}t  M{\"u}nchen, Scheinerstr. 1, D-81679 M{\"u}nchen, Germany\\
$^{9}$ Excellence Cluster ORIGINS, Boltzmannstr. 2, D-85748 Garching, Germany
}

\date{Accepted XXX. Received YYY; in original form ZZZ}

\pubyear{2022}

\begin{document}
\label{firstpage}
\pagerange{\pageref{firstpage}--\pageref{lastpage}}
\maketitle

\begin{abstract}
We study the evolution of dust in a cosmological volume using a hydrodynamical simulation in which the dust production is coupled with the MUPPI (MUlti Phase Particle Integrator) sub-resolution model of star formation and feedback. As for the latter, we keep as reference the model setup calibrated previously to match the general properties of Milky Way like galaxies in zoom-in simulations. However, we suggest that an increase of the star formation efficiency with the local dust-to-gas ratio would better
reproduce the observed evolution of the cosmic star formation density. Moreover, the paucity of quenched galaxies at low redshift demands a stronger role of AGN feedback.
We tune the parameters ruling direct dust production from evolved stars and accretion in the inter stellar medium to get scaling  relations involving dust, stellar mass and metallicity in good agreement with observations. In low mass galaxies the accretion process is inefficient. As a consequence, they remain poorer in silicate and small grains than higher mass ones.
We reproduce reasonably well the few available data on the radial distribution of dust outside the galactic region, supporting the assumption that the dust and gas dynamics are well coupled at galactic scales.
\end{abstract}




\section{Introduction}


Dust is a component of the interstellar matter (ISM) that,
besides affecting the radiation we receive from astrophysical objects, participates actively to galaxy evolution in many ways. It contributes to hot gas cooling: the collisions of ions with dust grains transfer to them thermal energy. This energy is then radiated in the infrared, and thus easily lost by the system \citep[e.g.][]{Burke1974,Dwek1981,Montier2004}. On the other hand, the metals subtracted from gas by grains would be effective ISM coolants while in gaseous form. The surfaces of grains catalyze, together with other reactions, the formation of H$_2$ molecules \citep[see review  by][]{Wakelam2017}. Therefore dust promotes the formation of molecular clouds, which are the sites of star formation, and it also shields them from stellar light.
The onset of galactic winds can be facilitated by dust, since its presence in the ISM enhances by orders of magnitude the action of radiation pressure \cite[e.g.][]{Murray2005}.
Dust can also promote the formation of a reservoir of low angular momentum gas in galaxies, providing accretion fuel to their super-massive black holes \citep[e.g.][]{Granato2004}.

For the above mentioned reasons, it is desirable as a first step to include a treatment of the complex processes responsible for dust production and evolution in galaxy formation models. We have now a broad understanding of the main processes affecting the dust content of galaxies. Seed grains condense in stellar ejecta, mostly Asymptotic Giant Branch stars (AGBs) winds and Supernovae (SNe) explosions. This primary grain population is then modified in the ISM by several processes. Gas particles can collide with the grain surface and {\it accrete} on it. This process affects the dust mass but also its overall chemical composition. However, if the collision is energetic enough, it can instead erode the grain. This {\it sputtering} process dominates in the hot plasma at  $T \gtrsim \mbox{a few} 10^5$ K. Accretion and sputtering are surface processes, and therefore they are most effective in small grains. Collisions between grains are also important, and can have opposite results affecting their size distribution. If the collision velocity is low enough, grains tend to {\it coagulate} to form bigger ones. This occurs in the densest regions of the ISM. Conversely, in diffuse ISM, grain-grain collisions are much faster, causing shattering into smaller pieces.

These admittedly complex and uncertain dust processes began to be incorporated into cosmological  galaxy formation computations only relatively recently, but still in a non-systematic way  \citep[e.g.][]{McKinnon2017,Aoyama2018,Gjergo2018,Vogelsberger2019,Hou2019,Li2019,Graziani2020,Granato2021}.
These works implemented ideas previously explored in the context of some one-zone computations \citep[e.g.][]{Dwek1998, Calura2008, Asano2013, Hirashita2015, Zhukovska2016} or non cosmological simulations \citep[e.g.][]{Bekki2015, Valentini2015, Aoyama2017}. Dust evolution processes have been incorporated also in some semianalytic models \citep[e.g.][]{Valiante2011,
Popping2017, Vijayan2019, Triani2020, Dayal2022}.
It is worth pointing out that even when the dust content of the ISM is estimated, its effects on the sub-resolution prescription for star formation and feedback are generally not considered. \\
In this paper, we analyze a cosmological box simulated with our most advanced star formation and feedback sub-resolution model MUPPI \citep[see][and references therein]{Murante2015,Valentini2020},  and including a state-of-the-art treatment of dust evolution, as described in \cite{Granato2021}. \rev{The present study
is the first one in which the evolution of the chemical composition
of dust and its size distribution are tracked simultaneously and independently for two dust species in a cosmological volume. Moreover, for the first time, we test in some detail the general performances of the most recent version of MUPPI, with included AGN physics, on cosmological volumes.} MUPPI has been developed, refined and calibrated in a series of papers on zoom-in cosmological simulations of Milky Way \revdue{(MW)} like galaxies. For this reason, in the present work, we generally take its parameters and setup as given, and we devote particular attention to the properties of the simulated galaxies hosted by dark matter (DM) halos featuring a mass within a factor $\sim 3$ from that of MW. In the following, we refer to these objects with the acronym MWHM (Milky Way Halo Mass) galaxies.
However, \rev{we also critically discuss the results of MUPPI when compared with observations related to the general galaxy populations, and in Section \ref{sec:SFRD+MS} we suggest possible modifications of the model to improve its performances. }
We adopt for the MW halo total mass the reference value $M_{200} \simeq 10^{12} \text{M}_\odot$\footnote{$M_{200}$ is the mass enclosed by a sphere whose mean density is 200 times the critical density at the considered redshift. The radius of this sphere is dubbed $R_{200}$.} \cite[for a recent compilation of estimates see][]{Callingham2019}.

After describing the simulations and recapping the main model features in Section \ref{sec:simulations}, we discuss how the simulated galaxy population matches the available observational constraints.
First (Sections \ref{sec:stellar_properties} and \ref{sec:mainpropMWgal}) we concentrate on general properties independent of dust.
Then we discuss the dust mass function of galaxies in Section \ref{sec:DMF}. Section \ref{sec:dustscaling} is devoted to scaling relations and involving the dust content. Then we focus in Section \ref{sec:dustoutgalaxies} on the dust that manages to exit from the galactic region.
Section \ref{sec:conclusions} presents a summary, some discussion and  future developments.

\section{Numerical Simulations}
\label{sec:simulations}
We performed cosmological hydrodynamical simulations with our branch of the GADGET3 code, which is a non-public evolution of GADGET2 \citep[][]{Springel2005}. The hydrodynamics is treated with the improved SPH (smoothed particle hydrodynamics) by \cite{Beck2016}.

We used the same cosmological parameters as the IllustrisTNG simulation  (https://www.tng-project.org), which in turn were selected in accordance with \cite{PlanckCollaborationXIII2016}. Thus we set
$\Omega_{\rm m}=\Omega_{\rm DM}+\Omega_{\rm b}=0.3089$, $\Omega_{\rm b}=0.0486$, $\Omega_\Lambda=0.6911$, $h=0.6774$, and adopted a power spectrum with primordial index $n_s=0.9667$ and normalization $\sigma_8=0.8159$.
Our main simulated  box has a size of $26 \, \text{cMpc}$ from initial conditions (ICs) obtained at $z=99$. To improve the statistic, and to test quickly the effects of model variations, we also run 4 boxes of size $13 \, \text{cMpc}$.
The initial conditions have been generated with the public code N-GenIC (https://www.h-its.org/2014/11/05/ngenic-code).

The mass resolution of DM particles is $M_\text{DM} \simeq 3.75 \cdot 10^7$~M$_\odot$, while the gas particles have initially a mass $M_\text{gas} \simeq 7 \cdot 10^6$~M$_\odot$. However the masses of gas particles undergo modifications due to star formation and chemical feedback (i.e.\ enriched gas restitution due to stellar evolution processes) from neighbouring stellar particles, which are spawned from the former ones during the evolution of the system.
The gravitational force is softened with a Plummer-equivalent softening length of 740 pc, constant in comoving units at $z>6$, and constant in physical units at lower redshift. \rev{The resolution adopted here is very similar to that of the AqC6 initial conditions, which have been extensively studied in \cite{Granato2021}, as well as in several previous MUPPI works not related to dust. However, as discussed in \cite{Murante2015}, MUPPI is weakly resolution dependent, as long as molecular cloud masses are not resolved \cite[see also Figure 13 in][]{Granato2021}.}

\subsection{Sub-resolution physics}
In the present work we adopted the same modelling as in \cite{Granato2021} for unresolved processes, namely
star formation, feedback, chemical enrichment and dust evolution. We additionally include the treatment of AGN feedback as in \cite{Valentini2020}.
The adopted IMF is that by \cite{Kroupa1993}.
We re-calibrated some parameters affecting {\it only} the dust properties, to improve the match with the available statistical properties of nearby galaxies (see Section \ref{sec:dtg}). However, these modifications do not affect significantly the results of that paper (see Section \ref{sec:dustmodel}).
While we refer the reader to \cite{Granato2021} for a full description of the model, we  provide below a  qualitative summary.

\subsubsection{Star formation, feedback, chemical enrichment}
\label{SFmuppi}
Star formation, stellar feedback and AGN feedback are processes that involve scales way below the resolution of cosmological simulations. We describe them with the model dubbed MUPPI (MUlti Phase Particle Integrator), first introduced by
\cite{Murante2010}, and later updated in various papers. In particular, here we adopted in general the version described in \cite{Valentini2020}, which implemented a treatment of AGN feedback. We also partly tested a couple of variations, as detailed in Section \ref{sec:stellar_properties}.

MUPPI aims at modelling phases of the ISM which are not captured by the numerical resolution. When a SPH gas particle becomes denser than $n_\text{thre} = 0.01 \, \text{cm}^{-3}$ and its temperature falls below $T_\text{thre}=50000\, \text{K}$, it is tagged as multiphase (MP), and it is treated as consisting of a hot gas phase, a cold gas phase and a virtual stellar component.  The two gaseous phases are assumed to be in pressure equilibrium. A set of ordinary differential equations describes the mass and energy flows between these components. These equations, numerically solved during the simulation, incorporate processes such as the radiative cooling of hot gas into cold one, the formation of stars from the molecular fraction of the latter and the evaporation of cold gas due to stellar and AGN feedback.
The star formation rate (SFR) of a multiphase particle is linked to the fraction $f_\text{mol}$ of cold gas which is in molecular form. This fraction is computed according to the prescription of \cite{Blitz2006}.
The chemical enrichment is accounted for as described in \cite{Tornatore2007}. Here we track the abundance of 15 elements (H, He, C, N, O, Ne, Na, Mg, Al, Si, S, Ar, Ca, Fe, and Ni) taking into account AGB stars, type Ia Supernovae (SNIa) and type II Supernovae (SNII).
We also include cooling due to dust grains, as described in Sec. 2.3.6 of \cite{Granato2021}.

\begin{figure*}
\centering
\includegraphics[width=\columnwidth]{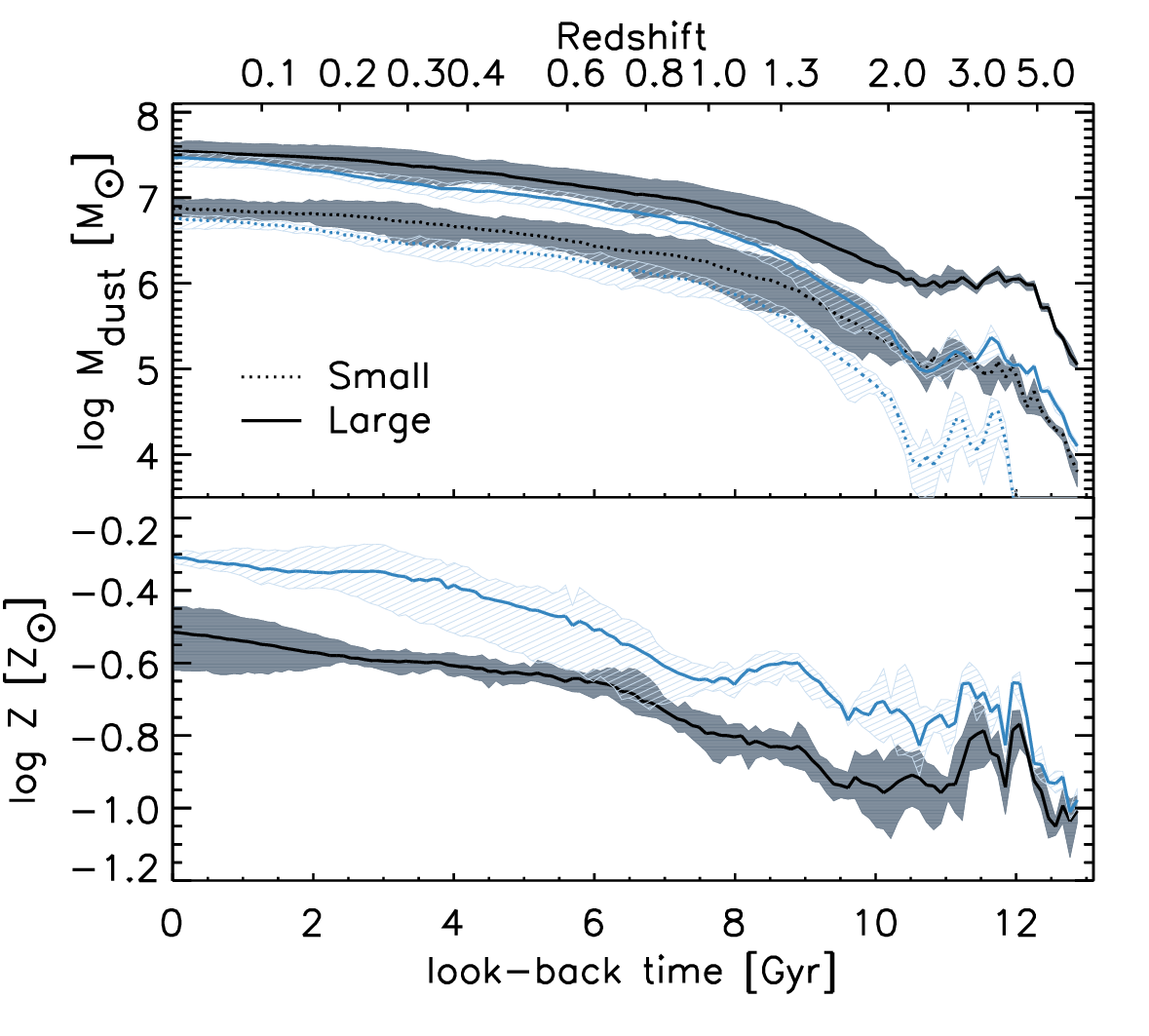}
\includegraphics[width=\columnwidth]{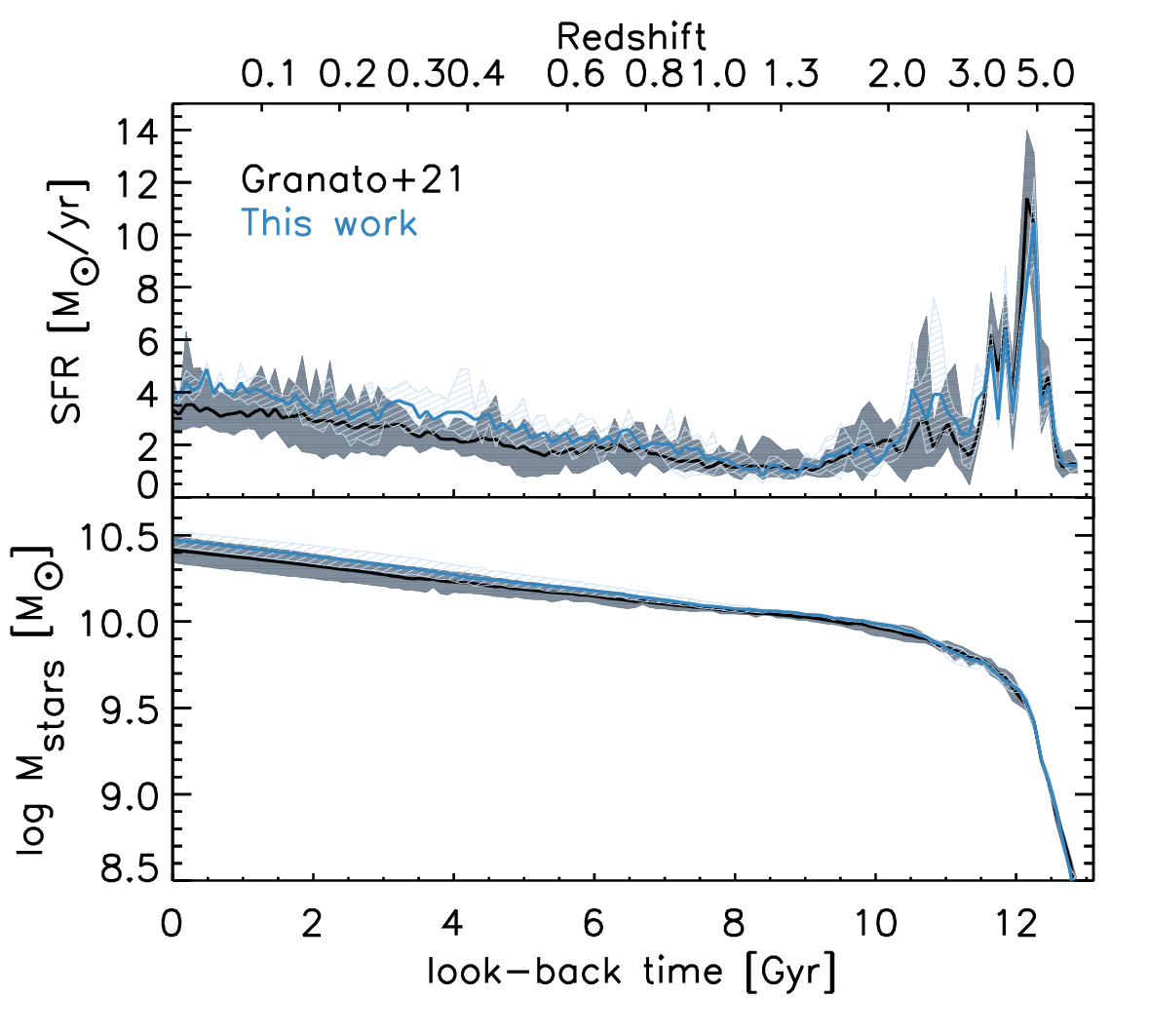}
\caption[width=\textwidth]{Comparison between the evolution of the AqC6 initial conditions used in \cite{Granato2021}, adopting the dust parameters from that work and those used in the present work, to improve the match with the statistical properties of nearby galaxies \rev{(see Sections \ref{sec:dustmodel} and \ref{sec:dtg})}. The panels show the time evolution of some fundamental quantities (dust mass \revdue{in Small and Large grains}, gas metallicity, \revdue{SFR} and stellar mass) of simulated disc galaxies. Both simulations have been repeated three times, introducing tiny perturbations in the initial conditions to quantify the effect of chaos (see Sec. 2.4.1 of \citealt{Granato2021}). The solid line and the shaded regions refer respectively to the median and the full dispersion of the three runs.}
\label{fig:gra21}
\end{figure*}

\subsubsection{Dust formation and evolution}
\label{sec:dustmodel}
We adopt the dust evolution model described in section 2.3 of  \cite{Granato2021}, to which we refer the reader for all the details.

We consider dust grains with different sizes and chemical composition. Specifically, we  follow the evolution of large and small grains according to the two size approximation by \cite{Hirashita2015}.
The approximation has been shown to produce results in good keeping with more complete, and numerically demanding, treatments of the size distribution \citep[see also][] {Aoyama2020}. We also differentiate between  carbonaceous and silicate grains according to their chemical composition. For the latter ones we adopt the Olivine composition MgFeSiO$_4$ \citep[see e.g.][]{Draine2003}. Thus we follow separately the abundances of four species of grains: large and small silicate ones, large and small carbonaceous ones.

Our treatment includes the various processes mentioned in the introduction, namely production of large grains by evolved stars (AGBs and SNae), accretion, sputtering, coagulation and shattering. Moreover, we compute hot gas cooling due to collisions of ions with grains. For most processes, we adopted the same efficiencies as in \cite{Granato2021}. However, in our fiducial model we decreased both the dust condensation efficiency in stars ejecta and the accretion efficiency, motivated by the observations discussed in Section \ref{sec:dtg}. As for the former efficiency, we adopted here $\delta_{AGB,sil} = \delta_{AGB,C} = 0.1$, $\delta_{SNII,sil}=\delta_{SNII,C}=0.1$.  These values are smaller by a factor three to five than those in \cite{Granato2021}, and similar to those adopted by other recent dust evolution models \citep[e.g.][]{Popping2017,Li2019}\footnote{see equations 2 to 6 in \cite{Granato2021}. These coefficients give the fraction of each metal entering dust that is {\it available} to be injected into the ISM as solid state grains rather than in gaseous form. Note, however, that, at variance with \cite{Popping2017}, \cite{Li2019} and several other papers, we adopt for silicate grains a treatment that always preserves the elemental mass partition of MgFeSiO$_4$. Therefore our actual condensation efficiencies are variable and smaller for these elements, but for the less abundant one. This element, dubbed {\it key element}, usually, but not always, turns out to be Silicon.
}. We also switched off the production of dust from SNIa \cite[as in][]{Li2019}, since recent works suggest that they are not significant sources of dust \cite[for a discussion and references see][]{Gioannini2017}. Moreover, in our model their contribution is well subdominant even when their condensation efficiency is maintained similar to that of the other two stellar channels.
As for the accretion efficiency, we decrease it by a factor of five with respect to \cite{Granato2021}. This variation can be thought of as a decrease of the effective sticking efficiency $S$ appearing in equation (11) of that paper. Alternatively, it could indicate that the fraction of ISM in which accretion is effectively at work, that is $F_\text{dense}$ in the same equation, is overestimated. This fraction is currently given by the fraction of multiphase particles that is molecular and star forming (see previous Section).  Another possibility is that the galaxies, as predicted by the fiducial model, are too gas-rich during their evolution. Indeed, the alternative model effDTG proposed in Section \ref{sec:SFRD}, featuring a lower gas fraction in galaxies, would demand a higher accretion efficiency to match the observations discussed in Section \ref{sec:dustscaling}. We will return to this point in that Section.

Although these variations with respect to \cite{Granato2021} are relatively significant, we verified that they do not affect substantially the results of that work, at least at $z \lesssim 1$. \rev{For example, Figure \ref{fig:gra21}. shows the evolution of stellar mass, \revdue{SFR}, dust mass, and gas metallicity for the disk-like galaxy produced by the AqC6 initial conditions, either adopting our re-tuned dust model or that used by \cite{Granato2021}. The differences in these and other quantities studied in the latter paper are modest at low z. \revdue{The most evident difference is in the gas metallicity, which is a factor $\sim1.5$ higher with the current set of parameters. This is essentially due to reduced condensation and grains accretion efficiency, resulting in a smaller overall depletion of gas metals.}} On the other hand, the same modifications are required to statistically reproduce  the scaling relations between gas metallicity and the dust-to-gas ratio derived for nearby galaxies, as we will show in Section \ref{sec:dtg}.


\subsection{Galaxies identification}
The Friends-of-Friends (FoF) algorithm is used to arrange particles in groups with a linking length of 0.16 the mean interparticle separation. Then the gravitationally bound sub-structures of the  identified halos are detected by applying the Subfind algorithm \citep{Springel2001, Dolag2009}. Throughout this paper we restrict our analysis to central galaxies, which are associated with the main subhalo of a FoF halo, i.e. the subhalo which hosts the most gravitationally bound particle of the group. \\
The quantities of each galaxy (e.g. stellar mass, metallicity, dust mass) are computed considering gas and stellar particles within a 3D aperture with radius $R_\text{gal}=0.1 R_{200}$, including only particles belonging to the corresponding subhalo. We will generally refer to $R_\text{gal}$ as the radius of the galaxy. In our analysis we consider only galaxies with stellar mass $M_\text{stars} \gtrsim 10^8 \, M_\odot$, which corresponds to $\gtrsim 100$ stellar particles. \\

%
%

\section{Results}

\subsection{Stellar properties}
\label{sec:stellar_properties}

\begin{figure*}
    \centering
    \begin{subfigure}[t]{0.3\textwidth}
        \centering
        \includegraphics[width=\linewidth]{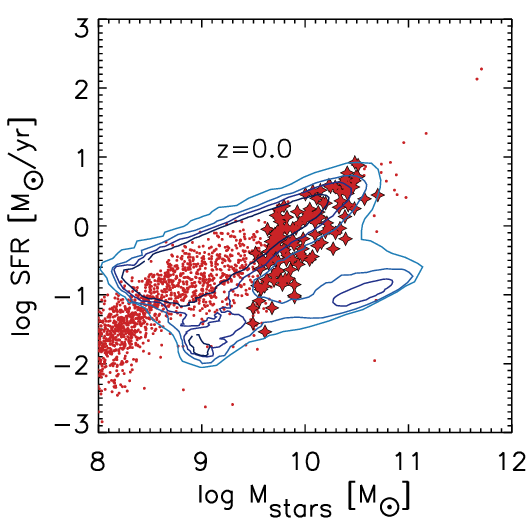}

    \end{subfigure}
    \hfill
    \begin{subfigure}[t]{0.3\textwidth}
        \centering
        \includegraphics[width=\linewidth]{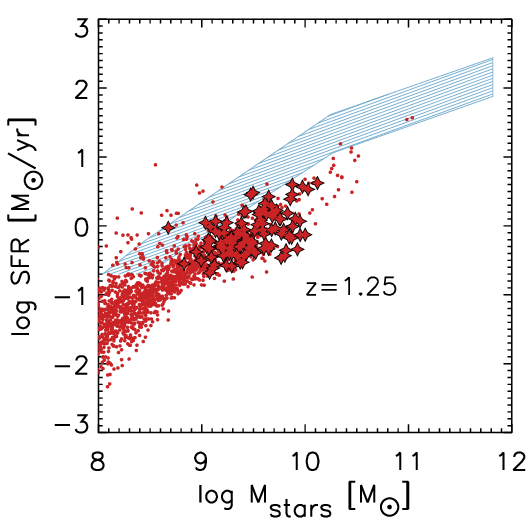}

    \end{subfigure}
    \hfill
    \begin{subfigure}[t]{0.3\textwidth}
        \centering
        \includegraphics[width=\linewidth]{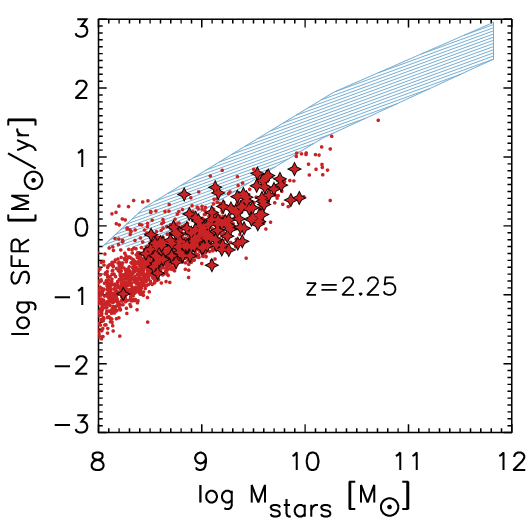}

    \end{subfigure}

    \caption{Main Sequence of the set of fiducial runs at different redshifts.   Star symbols refer to MWHM galaxies and their progenitors. At $z=0.0$ we compare our results with \citealt{Renzini2015}.
    \rev{Four contour levels from their Figure 4 are shown. From lighter to darker blue they correspond to number density of
    $\simeq 4,8,11,13 \cdot 10^4$. At higher redshift we show with shaded regions the distribution of Main Sequence galaxies from \citealt{Leja2019}.}}
    \label{fig:mainsequence}

\end{figure*}

\begin{figure*}
    \centering
    \begin{subfigure}[t]{0.3\textwidth}
        \centering
        \includegraphics[width=\linewidth]{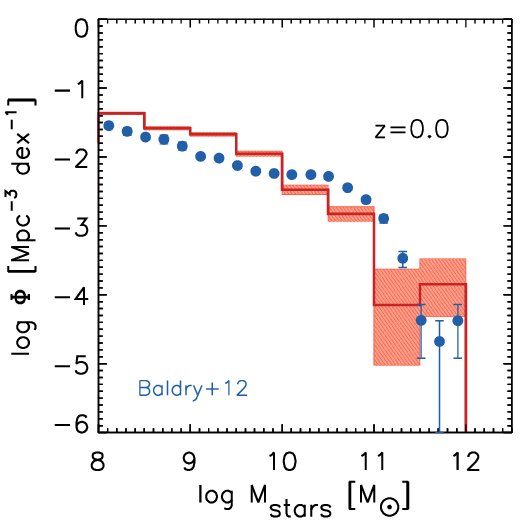}
    \end{subfigure}
    \hfill
    \begin{subfigure}[t]{0.3\textwidth}
        \centering
        \includegraphics[width=\linewidth]{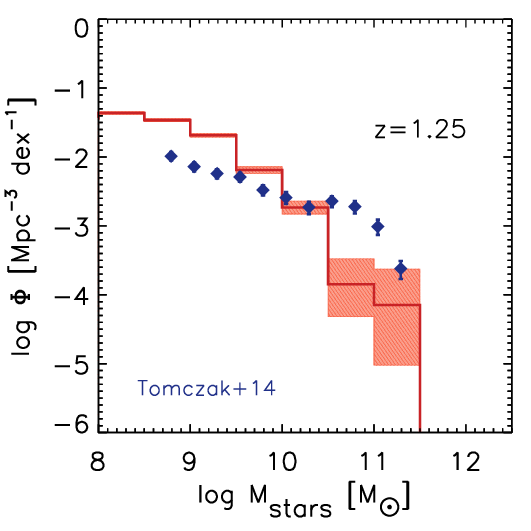}
    \end{subfigure}
    \hfill
    \begin{subfigure}[t]{0.3\textwidth}
        \centering
        \includegraphics[width=\linewidth]{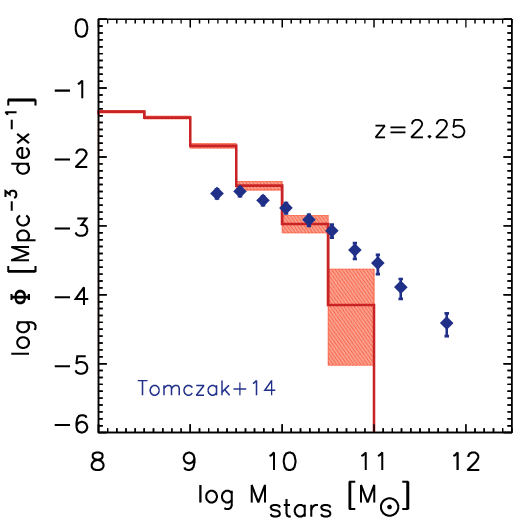}
    \end{subfigure}

    \caption{Stellar Mass Function of the set of fiducial runs at different redshift; the shaded area is the Poissonian error. The local determination of \citet{Baldry2012} is shown, while we compare with \citet{Tomczak2014} data in the $z=1.25$ and $z=2.25$ panels. }
    \label{fig:stellarMF_z}
\end{figure*}

\begin{figure}
\centering
\includegraphics[width=1.\columnwidth]{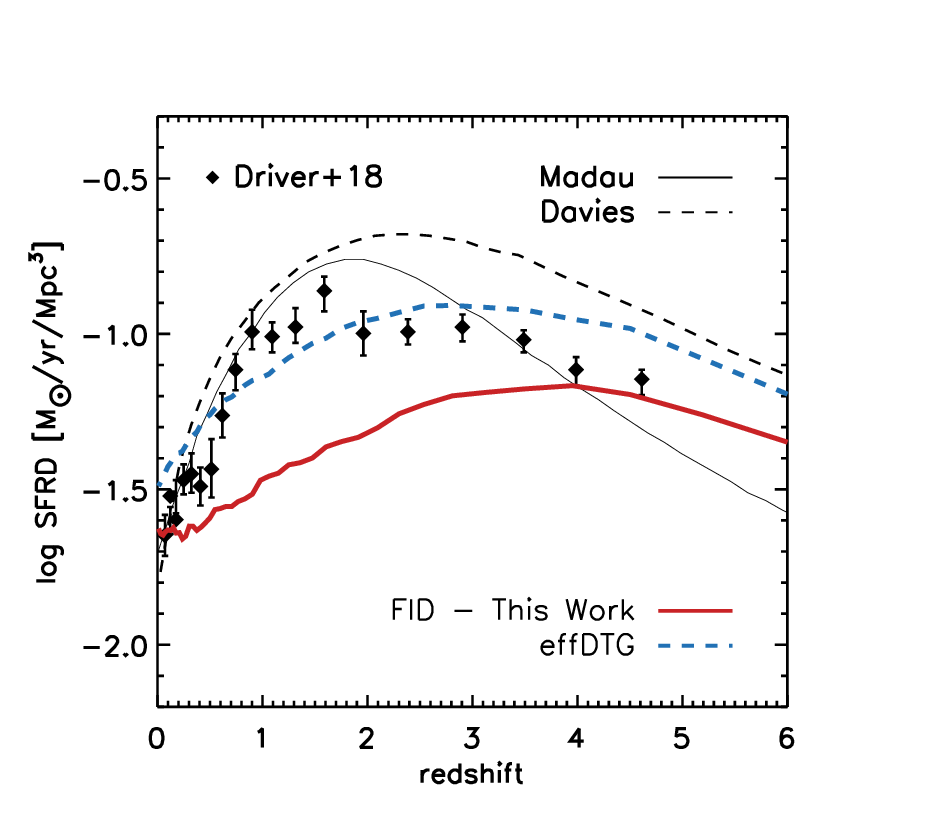}
\caption[width=\textwidth]{Median Star Formation Rate Density of the boxes used in this work (FID; \textit{red}). We also show a possible improvement of our results discussed in Section \ref{sec:SFRD} (effDTG; \textit{blue}). We over-plot some SFRD(z) determinations from literature: the \textit{solid} and \textit{dashed} black lines show the fits to the cosmic SFRD obtained by \cite{Madau2014} and \cite{Davies2016}, while the \textit{black points} represent the determination of \cite{Driver2018}. All these data have been converted to be consistent with our IMF \citep[][]{Kroupa1993} by exploiting the coefficients in Table 1 of \cite{Driver2013}.}
\label{fig:SFRD}
\end{figure}

This section discusses how our fiducial simulation compares with a few basic observational quantities related to the stellar component of the entire galaxy population. However, we recall that the purpose of the present work is to test our modelling of dust evolution on a relatively large sample of simulated galaxies, focusing on those having a mass comparable to that of the MW, on which MUPPI has been previously tailored, rather than to adjust MUPPI to match the general galaxy population in the box.
Unsurprisingly, it turns out that it would require some model adjustments to achieve the latter result.

\label{sec:SFRD+MS}

\subsubsection{Galaxy Main Sequence}
It is well known that galaxies concentrate in well defined regions in the star formation - stellar mass plane \cite[see][and references therein]{Renzini2015}.
In particular, star forming galaxies define the so called main sequence (MS), since the two quantities turns out to be strongly correlated. The MS is the most populated region of the plane in number, while the mass budget in the local Universe is dominated by quenched galaxies, featuring a SFR about an order of magnitude below the MS.

In Figure \ref{fig:mainsequence},
we show the distribution of simulated galaxies in the plane $M_{\rm stars}$-SFR in three redshift snapshots, together with a selection of recent determinations.
At $z=0$ we over-plot \rev{four} density contours taken from Figure 4 by \cite{Renzini2015}. Their spectroscopic sample is extracted from the Sloan Digital Sky Survey DR7 release and includes $\sim 240,000$ galaxies.
At $z=1.25$ and $2.25$ we show the recent observational determinations of the \revdue{MS} of star-forming galaxies by \cite{Leja2019} in the redshift ranges $1.0<z<1.5$ and $2.0<z<2.5$ respectively.  To properly compare our results with those derived from observations, we convert the observational estimates of SFRs and stellar masses to our adopted IMF \citep{Kroupa1993} \footnote{As for SFRs, we use the coefficients reported Table 1 of \cite{Driver2013}. The conversion of stellar masses is more uncertain since it depends on the region of the SED used to estimate the stellar mass and on the star formation and metal enrichment histories. Throughout this work, when we need to compare our predictions with observed stellar masses, we exploit the results by \cite{Portinari2004}, who provides the $M/L_\text{I band}$ ratios for different IMFs. Namely, we scale the observed stellar masses by a factor $(M/L_\text{I,K93}) / (M/L_\text{I,IMF})$, where the subscripts K93 and IMF refer respectively to the \cite{Kroupa1993} IMF (adopted in this work) and to the generic IMF adopted to derive the observed stellar masses. For the metallicity values investigated by \cite{Portinari2004}, the correction turns out to be $\simeq 1.1$ and and $\simeq 1.4$ for a \cite{Chabrier2003} and \cite{Salpeter1955} IMFs respectively}. Figure \ref{fig:mainsequence} displays a reasonable qualitative agreement at $z=0$ between the distribution of our simulated galaxies and observed \revdue{MS} in the $M_{\rm starst}$-SFR plane. However, our simulation features a dearth of quenched galaxies, at least at $z=0$. We verified this problem could be allieviated by increasing the SMBH growth and related AGN feedback, with respect to the setup adopted in \cite{Valentini2020} (see Appendix \ref{app:morebh}).
At the same time, this modification yields a correlation between the SMBH and spheroidal stellar component in better agreement with the observations. However, since the results discussed here on dust are largely unaffected, we stick to the setup of \cite{Valentini2020} for the fiducial run. \\
At higher redshift (central and right panels of Figure \ref{fig:mainsequence}), the simulated galaxies display a deficiency of star formation activity by a factor $\sim 3$ at a given stellar mass, in keeping with the under-prediction of the SFRD discussed in Section \ref{sec:SFRD}. Indeed, the alternative model effDTG discussed there would be in somewhat better agreement with the observations at those redshifts.

\subsubsection{Stellar Mass Function}
\label{sec:SMF}
The Stellar Mass Functions (SMFs) predicted by the simulation at the same three cosmic epochs are shown in Fig. \ref{fig:stellarMF_z}, together with a few observational estimates.  At $z=0$, the agreement is reasonable, with a moderate tendency for an overproduction of galaxies at $M_\text{stars} < 10^{10} \text{M}_\odot$ and an underproduction above this limit. The problem worsens at higher redshift. However, at high mass the statistical significance is marginal due to the size of the simulated sample. We point out that, in general, state of the art cosmological simulations require, to get a good match with observed SMFs, complex, and to some extent ad-hoc, scaling of galactic winds velocity and/or efficiency with galaxy properties (e.g.\ metallicity and mass) and even with redshift \citep[e.g.][]{Pillepich2018}.
By converse, in the present simulation,  the fraction of SNae energy going into kinetic winds (the {\it wind efficiency}) is simply constant, and the velocities of wind gas particles are computed consistently from the energy given to each of them from SNae explosions occurring in nearby stellar particles \citep[][]{Valentini2017}.

\subsubsection{Cosmic Star Formation Rate}
\label{sec:SFRD}
Figure \ref{fig:SFRD} shows the redshift evolution of the total cosmic \rev{Star Formation Rate Density} (SFRD) of the simulated box. We over-plot the observational determinations of the cosmic SFRD by \cite{Madau2014}, \cite{Davies2016} and \cite{Driver2018}. Unfortunately, the SFRD predicted by our fiducial run (dubbed FID) is quite at odd with observational estimates, featuring a broad peak at too high $z\sim 4$, and too low values at lower redshift. The largest discrepance is at redshift $\sim 2-3$, where we under-predict the SFRD by a factor $\sim 3$.
It is worth pointing out that numerous studies have been devoted to check the consistency between the histories of the SFRD and the cosmic stellar mass density (SMD). In many cases, the general result has been that the former should be revised downward to match the latter, especially around $z \sim 2$. This discrepancy could be due to several reasons that could lead to an underestimate of the SMD or an overestimate of the SFRD \cite[see][and references therein]{Yu2016}. In the latter case, the SFRD would better agree with our fiducial simulation. As a matter of fact, we have seen in Section \ref{sec:SMF} that the predicted \revdue{SMF} in the local universe matches reasonably well the observations. As a consequence, the final stellar mass density turns out to be well reproduced. Indeed, the integral of the observed and simulated SMFs shown in the left panel of Figure \ref{fig:stellarMF_z} differs by a modest $\sim 0.1$ dex.

We devoted some effort to investigating a model improvement on this aspect, made possible by treating the dust content of the ISM. Indeed, a better match with the SFRD can be obtained if the \rev{star formation} efficiency tends to increase with cosmic time. A physically plausible idea is to assume that it increases with the dust-to-gas (DTG) ratio of the particle. It is conceivable that a dust enriched ISM favours the process of star formation via several mechanisms not captured at resolved levels.
\rev{Star formation occurs in cold, dense, dusty molecular clouds, whose masses and sizes are far below our resolution. In these regions, dust plays a crucial role in determining the amount of star-forming gas, via processes such as promoting the formation of H$_2$ molecules or shielding gas from the UV interstellar radiation field. It is well established that whenever dust is present in the ISM, catalytic reactions on surfaces of interstellar grains become the main channel of H$_2$ formation \citep[for a recent review see][]{Wakelam2017}. Moreover, it has been suggested that dust shielding could be even more important than the presence of H$_2$ in enabling star formation in gas clouds \citep[e.g.][]{Krumholz2011,Glover12,Byrne2019}. Higher DTG molecular clouds may thus be able to convert a higher fraction of their mass into stars before being destroyed, thanks to their higher shielding and enhanced capability to produce new H$_2$ molecules. Inspired by these considerations, we decided to test a dependence of the star formation efficiency on the DTG\footnote{We exploit here our dust model and make use of the DTG of the SPH gas particles hosting star formation. However, due to the DTG-Z correlation (see Sec. \ref{sec:dustscaling}), a Z-dependent star formation efficiency may also work in the same direction.}, as detailed below.}

We show in the Figure also the model effDTG where the \revdue{SFR} is given by
\begin{equation}
\label{newSF}
    \Dot{M}_\text{sf} = f_*(\text{DTG}) \cdot \frac{f_\text{mol}M_\text{c}}{t_\text{dyn,c}}.
\end{equation}
In the above formula, $M_\text{c}$ is the cold mass of the MP particle, $f_\text{mol}$ its molecular fraction, and $t_\text{dyn,c}$ is the cold gas dynamical time.
The factor $f_*(\text{DTG})$ represents the star formation efficiency of the molecular clouds. In previous MUPPI works, and in the fiducial model here, \rev{the same equation is used, but with a constant and quite low value for $f_*=0.02$}. The model effDTG adopts instead the following dependence on the \revdue{DTG} ratio of the particle:
\begin{equation}
    \label{fstarDSG}
    f_* (\text{DTG}) = \left(\alpha + \frac{\beta-\alpha}{1 + \left(\frac{\text{DTG}}{\text{DTG}_\text{crit}}\right)^\gamma}\right)^{-1}.
\end{equation}
We found that a good match with the observed SFRD can be obtained by setting $\beta=50$, $\alpha=12$, $\text{DTG}_\text{crit}=2\cdot 10^{-5}$ and $\gamma = 0.8$. With these values, the star formation efficiency increases smoothly with DTG, ranging from $0.02$, when  there is no dust, to $\simeq 0.08$, when $\text{DTG} >> \text{DTG}_\text{crit}$. \rev{We remark that Eq. \ref{fstarDSG} simply aims at describing an increasing star formation efficiency with DTG, with a smooth functional form covering reasonable star formation efficiency values (e.g. \citealt{Kim21})}

Despite the encouraging result shown in Figure \ref{fig:SFRD}, we do not adopt this solution in the rest of the paper.  We verified that it would require more tuning of  MUPPI parameters to reproduce {\it in details} the good results shown in the previous work using MUPPI in zoom-in simulations of MW like galaxies.
Such as study is clearly outside the scope of the present work, and will be  the subject of future investigation. Therefore we maintain here the setup that has been carefully studied previously. \rev{However, to assess the robustness of our findings, in Appendix \ref{app:effdsg} we compare our fiducial results to the ones obtained when adopting the model with enhanced star formation.}

To sum up the results of this Section \ref{sec:stellar_properties}, our present model has troubles in reproducing the cosmic SFRD.
Simulated galaxies distribute satisfactorily in the plane $M_{\rm stars}$-SFR, at least as far as the \revdue{MS} is concerned. The local SMF is in acceptable agreement with available data, more so for galaxies whose stellar mass lies within a factor $\sim10$ that of the MW. This finding is in keeping with the fact that  MUPPI has been calibrated on this galaxy mass scale. Therefore, in the rest of the paper we deserve particular attention on galaxies that forms in halos whose final mass $M_{200}$ lies in the interval from $3 \times 10^{11}$ to $3 \times 10^{12}$ $\text{M}_\odot$, that is one dex centered on the  estimated mass of the MW halo $10^{12} \text{M}_\odot$ \cite[see][and references therein]{Callingham2019}. For these galaxies (henceforth MWHM galaxies) we presented good reasons to believe that MUPPI captures the main features of the unresolved processes of star formation and feedback, and we give a few more reasons in the next Section. In the remaining figures, we highlight them with larger symbols than those used for the general simulated population. \rev{The same large symbols are used at $z>0$ for MWHM progenitors, which are identified by means of the most massive progenitor halo in the merger tree.}

\begin{figure*}
\centering
\includegraphics[width=0.8\textwidth]{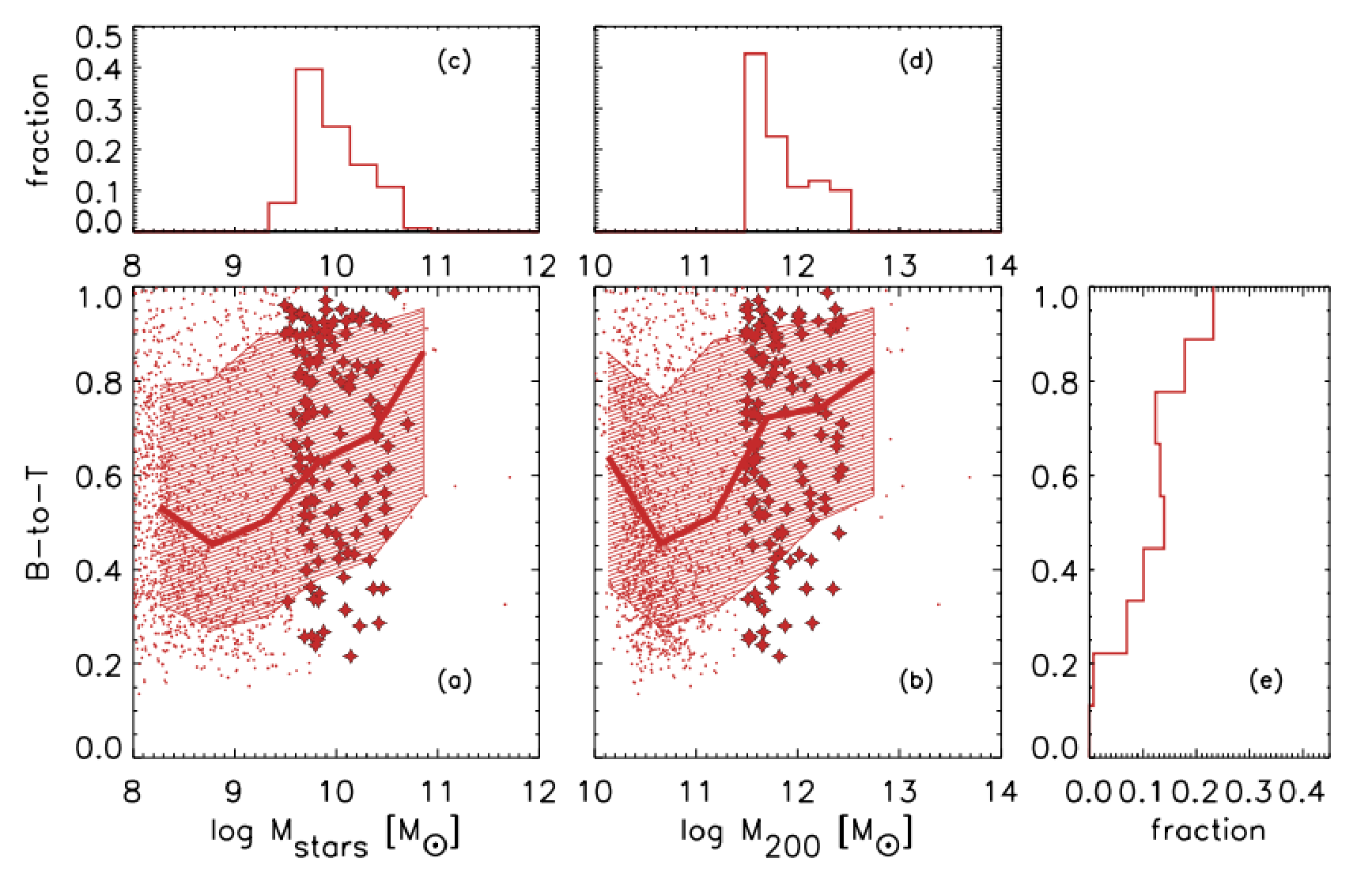}
\caption[width=\textwidth]{Bulge-to-Total ratio as a function of the stellar mass (a) and halo mass (b) \rev{at $z=0.0$}. We show the full sample of our simulated galaxies, and mark MWHM galaxies with a star symbol. In small panels we show the distribution of $M_\text{stars}$ (c), $M_\text{200}$ (d) and B-to-T (e) restricted to MWHM galaxies. The solid line marks the median of all galaxies, while the shaded region encloses the 25-75th percentiles.}
\label{fig:BtoT_Ms_M200}
\end{figure*}

\begin{figure*}
\centering
\includegraphics[width=0.8\textwidth]{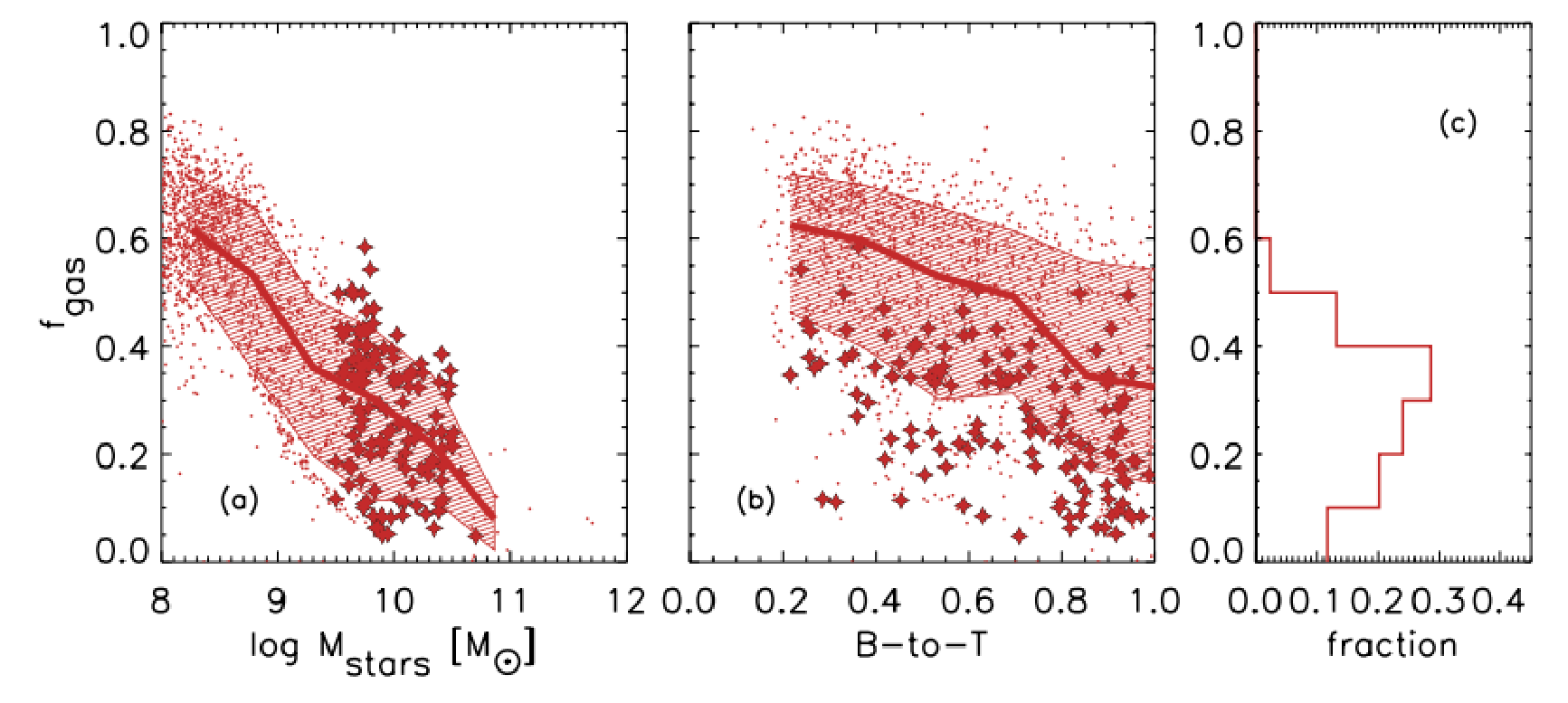}
\caption[width=\textwidth]{Gas fraction $f_\text{gas} = M_\text{gas,MP}/(M_\text{gas,MP}+M_\text{stars})$ as a function of the stellar mass (a) and Bulge-to-Total ratio (b) \rev{at $z=0.0$}. We show the full sample of our simulated galaxies, and mark MWHM galaxies with a star symbol. In the right small panel (c) the distribution of $f_\text{gas}$ of the MWHM galaxies is shown. The solid line marks the median of all galaxies, while the shaded region encloses the 25-75th percentiles.}
\label{fig:fgas_Ms_M200}
\end{figure*}

\begin{figure*}
\centering
\includegraphics[width=0.8\textwidth]{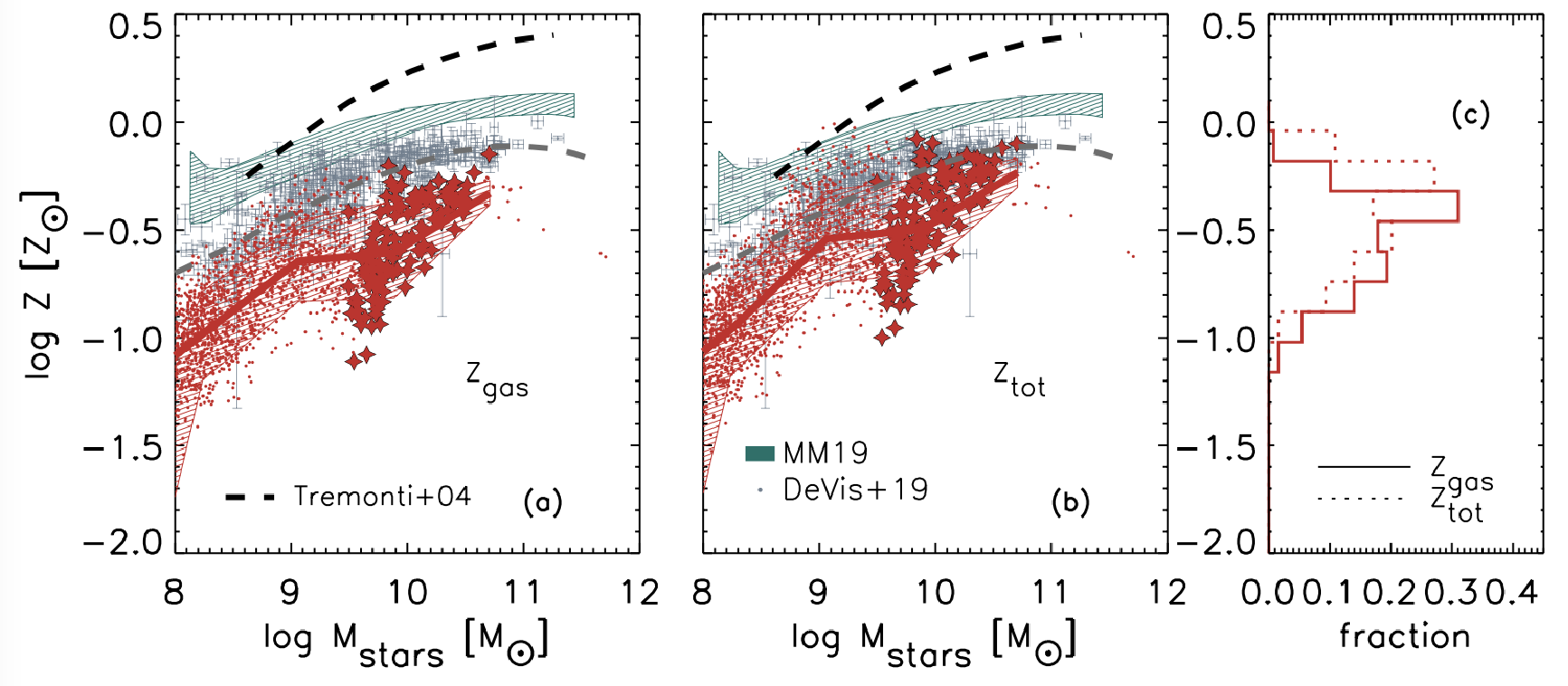}
\caption[width=\textwidth]{Mass-Metallicity relation for all our simulated galaxies \rev{at $z=0.0$} (MWHM galaxies are marked with a star symbol). We show both the gas metallicity ($Z_\text{gas} = M_\text{metals,gas} / M_\text{gas}$, panel a) and the total ISM metallicity $Z_\text{tot} = (M_\text{metals,gas}+M_\text{metals,dust}) / M_\text{gas}$, panel b), e.g. including also the metals locked in dust grains. Some determinations from literature are shown \citep{Tremonti2004, Maiolino2019, Vis2019}. \cite{Vis2019} data refer to the DUSTPEDIA sample, with their reference (PG16S, their Section 3.2) calibration for metallicites and stellar masses derived by \cite{Nersesian2019}. In panel (c) the distribution of both $Z_\text{gas}$ and $Z_\text{tot}$ is shown for MWHM galaxies. The solid line marks the median of all galaxies, while the shaded region encloses the 25-75th percentiles.}
\label{fig:Zgas_Ms_M200}
\end{figure*}

\subsection{Main properties of MWHM galaxies}
\label{sec:mainpropMWgal}
In this Section we discuss some properties, not directly related to dust, of the simulated galaxies formed in DM halos whose mass is within a factor $\sim 3$ that of the MW (MWHM galaxies).

Figure \ref{fig:BtoT_Ms_M200} shows 1D and 2D distributions of the bulge over total mass ratio B/T as a function of stellar and halo mass.
We estimate B/T from the star particles 6D phase space information, \rev{following \cite{Murante2015}}. We compute the orbital circularity $\epsilon_j=J_{\hat{z},j}/J_{circ,j}$ of stellar particles within $R_\text{gal}$. Here, $\hat{z}$ is the net spin axis of the galaxy, $J_{\hat{z},j}$ is the z component of the particle $j$ specific angular momentum and $J_{circ,j}$ is its specific angular momentum if the particle had a circular orbit with the same energy. To obtain the bulge mass, we sum the mass of the stellar particles with $\epsilon_j < 0$, and multiply it by 2, \rev{under the hypothesis that the bulge is supported by velocity dispersion and the number of co- and counter-rotating stars is equal.}

The resulting distribution of B/T for MWHM galaxies is quite broad, featuring a median of $0.68$ and an inter-quartile range $0.47$-$0.83$.
This finding is broadly consistent with observational results and with those from other simulations \citep[see for instance figure 3 in][and references therein]{Tacchella2019}. However, the details depend on the different methods used to derive B/T, \rev{as discussed in section 3.2 of the latter paper}.

Figure \ref{fig:fgas_Ms_M200} shows 1D and 2D  distributions of the gas fraction  $f_{\rm gas}=M_{\rm gas,MP}/(M_{\rm gas,MP}+M_{\rm stars})$\footnote{The subscript MP refers to multiphase gas. We compute this fraction excluding hot non multiphase particles since most observations refers to the cold gas content of galaxies. MP particles are dominated in mass by its cold component.}
as a function of the stellar mass and of B/T. The median gas fraction  of MWHM galaxies is 0.3, with a tendency for lower values at larger B/T, as expected. In particular "later type galaxies" feature $f_{\rm gas}$ values somewhat larger than those found in nearby late type galaxies \citep[e.g.][]{Saintonge2022}.

Figure \ref{fig:Zgas_Ms_M200} displays 1D and 2D  distributions of, respectively, gas phase metallicity $Z_\text{gas}$ and total ISM metallicity $Z_\text{tot} = (M_\text{metals,gas}+M_\text{metals,dust}) / M_\text{gas}$ \footnote{In this work we assume $Z_\odot = 0.0134$ following \citet{Asplund2009}.}.
We reproduce reasonably well the slope and the dispersion of the stellar mass-metallicity relations. As for its normalization, which is still uncertain by a factor $\sim 3$ \citep{Maiolino2019}, we are possibly low by a factor $\gtrsim 2$. It is worth pointing out that most previous model determinations not including dust evolution, neglected the fact that a significant fraction of metals $\gtrsim 50 \%$ is depleted to dust (e.g.\ Figure \ref{fig:mdZ}), while this is not the case in the present work. To better illustrate this point, the central panel of this figure shows $Z_\text{tot}$.

In this Section we showed that MWHM galaxies in our boxes have a "morphological" distribution in reasonable agreement with observations, a possibly somewhat high gas fractions, and metallicities likely low by a factor $\sim 2$. The latter two problems may be partly related to an excessive presence of scarcely processed gas in the galactic region. However, other possible ways out from the low metallicity issue are a more top-heavy IMF, or more generous stellar yields\footnote{For a detailed study of the effects of different IMFs and stellar yields in zoom-in simulations of MW like galaxies performed with MUPPI see \cite{Valentini2019}}. Finally, we point out that the stellar masses produced by the selected halos are low by a factor about $3$. Indeed (see Figure \ref{fig:BtoT_Ms_M200}) even halos whose total mass is three times that of the MW, produce galaxies whose stellar mass is similar to that estimated for the MW $\simeq 5 \times 10^{10} M_\odot$ by \cite{Cautun2020}. However, this problem is largely inherited from the adopted MUPPI setup, from \cite{Valentini2020}, which result in a galaxy whose mass is about 60\% that of the MW (see table 5 in the latter work) in a DM halo \citep[dubbed AqC,][]{Springel2008} almost twice as massive as the MW halo $M_{200} \sim 1.8 \times 10^{12}$ M$_\odot$.


\subsection{Dust Mass Function}
\label{sec:DMF}
\begin{figure*}
    \centering
    \begin{subfigure}[t]{0.3\textwidth}
        \centering
        \includegraphics[width=\linewidth]{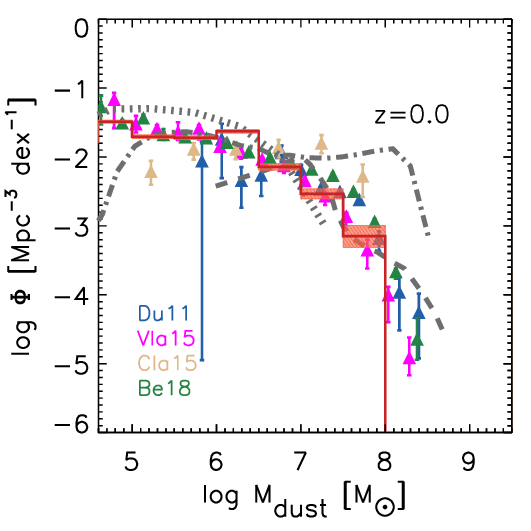}
    \end{subfigure}
    \hfill
    \begin{subfigure}[t]{0.3\textwidth}
        \centering
        \includegraphics[width=\linewidth]{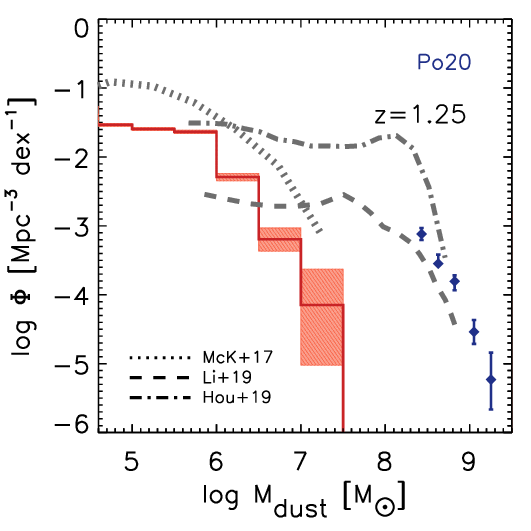}
    \end{subfigure}
    \hfill
    \begin{subfigure}[t]{0.3\textwidth}
        \centering
        \includegraphics[width=\linewidth]{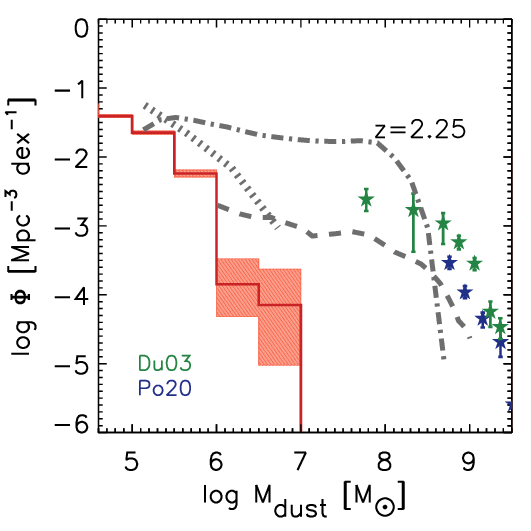}
    \end{subfigure}

    \caption{Dust Mass Function of the set of fiducial runs at different redshifts; the shaded area is the Poissonian error. We show observations from \citet{Dunne2011}, \citet{Vlahakis2005}, \citet{Clark2015}, \citet{Beeston2018} ($z=0$), \citet{Pozzi2020} ($z=1.25$ and $z=2.25$) and \citet{Dunne2003} ($z=2.25$). \rev{Results from other cosmological simulations including dust at similar redshifts are shown as well: \citet{McKinnon2017} ($z=0.0, \, z=1.0,\, z=2.5$), \citet{Li2019} ($z=0.0, \, z=1.0,\, z=2.0$), and \citet{Hou2019} ($z=0.0, \, z=1.0,\, z=2.0$)
    .}}
    \label{fig:dustMF_z}
\end{figure*}

Figure \ref{fig:dustMF_z} displays the simulated Dust Mass Function (DMF). Whereas the local observations are in excellent agreement with our simulation results, we heavily under-predict the abundance of dust rich galaxies ($M_\text{dust} \gtrsim 10^8 \, M_\odot$) at $z=1.25$ and $2.25$.
A similar problem has been found by \cite{McKinnon2017} (when calibrated to match the local DMF) and, to a lesser extent, by \cite{Li2019}, while the simulations by \cite{Hou2019} over-predict the number of very dusty system at $z=0$, and under-predict them at $z=2.5$. \rev{In conclusion, most current cosmological simulations do not reproduce well the observed DMF over cosmic time.\\}
\revdue{Identifying the reasons behind the differences between cosmological simulations is complex and outside the scope of the present work. We only point out that, while the dust processes included are the same,
their implementation differs in several aspects. For instance, none of the other simulations follows at the same time two grain sizes and two different chemical compositions, and only that by \cite{Hou2019} treats the size distribution, adopting the same two-size approximation we use. Moreover, our simulation is the only one in which the accretion process maintains a given element proportion for silicates. Furthermore, the various simulations adopt different sub-resolution models for baryonic physics.}


\subsection{Dust scaling relations}
\label{sec:dustscaling}

In this section we focus on the relationship between the dust content of simulated galaxies and other properties, mostly at $z=0$.

\begin{table}

\centering


 \begin{tabular}{c c c c c c}
 \hline
 \multirow{2}{*}{Run} & \multicolumn{3}{c}{Condensation Efficiency} & Accretion \\ [0.5ex]
 & $\delta_\text{SNII}$ & $\delta_\text{SNIa}$ & $\delta_\text{AGB}$ & $\tau_\text{acc}$  \\
\hline\hline
condG21-accG21 & $0.8$ & $0.8$ & $1.0$ & $\tau_\text{G21}$  \\
condG21-acc0.2 & $0.8$ & $0.8$ & $1.0$ & $5 \cdot \tau_\text{G21}$  \\
COND0.1-accG21 & $0.1$ & $-$ & $0.1$ & $\tau_\text{G21}$  \\
COND0.1-acc0.2 & \multirow{2}{*}{$0.1$} & \multirow{2}{*}{$-$} & \multirow{2}{*}{$0.1$} & \multirow{2}{*}{$5 \cdot \tau_\text{G21}$}  \\
(fiducial this work) & & & & \\
 \hline
 \end{tabular}
\caption{Dust parameters adopted in runs of Fig. \ref{fig:mdZ_models}. The condensation efficiencies for all the stellar channels are shown (they are the same for both carbonaceous and silicate grains), as well as the accretion timescale with respect to \citet{Granato2021} choices. These latter parameters are introduced respectively in Eq. 11 and 14 of \citet{Granato2021}.}
\label{tab:DustModels}
\end{table}


\subsubsection{Dust, Stellar mass and SFR}

\begin{figure}
\centering
\includegraphics[width=0.5\textwidth]{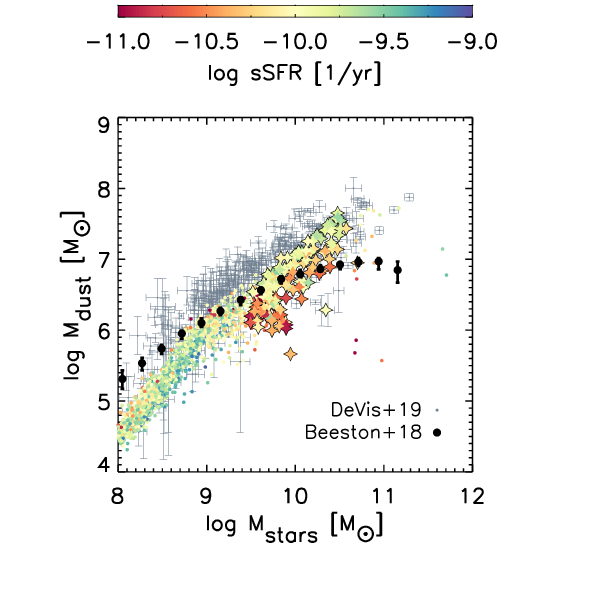}
\caption[width=\textwidth]{Dust mass as a function of the stellar mass of our simulated galaxies at $z=0$, while the specific SFR is shown as color code. Star symbols are the MWM galaxies in our sample. Data from \cite{Vis2019}, \cite{Beeston2018}. \cite{Vis2019} data refer to the DUSTPEDIA sample, with their reference (PG16S, their Section 3.2) calibration for metallicites and stellar masses derived by \cite{Nersesian2019}.}
\label{fig:msmd}
\end{figure}

Figure \ref{fig:msmd} shows how the dust mass $M_\text{dust}$  correlates with the stellar mass $M_\text{stars}$, with colors coding the sSFR of each simulated galaxy. We compare our results with various observations. The dust content of galaxies increases almost linearly with the stellar mass, and, quite unsurprisingly, galaxies with higher sSFR at a given stellar mass contains more dust.
Similarly, observations display a monotonic increase of dust mass with stellar mass.
The determinations presented recently in \cite{Vis2019} feature a linear increase of the dust mass with the stellar mass, similar to that of our simulations, while older estimates suggest a milder growth.

\begin{figure}
\centering
\includegraphics[width=0.4\textwidth]{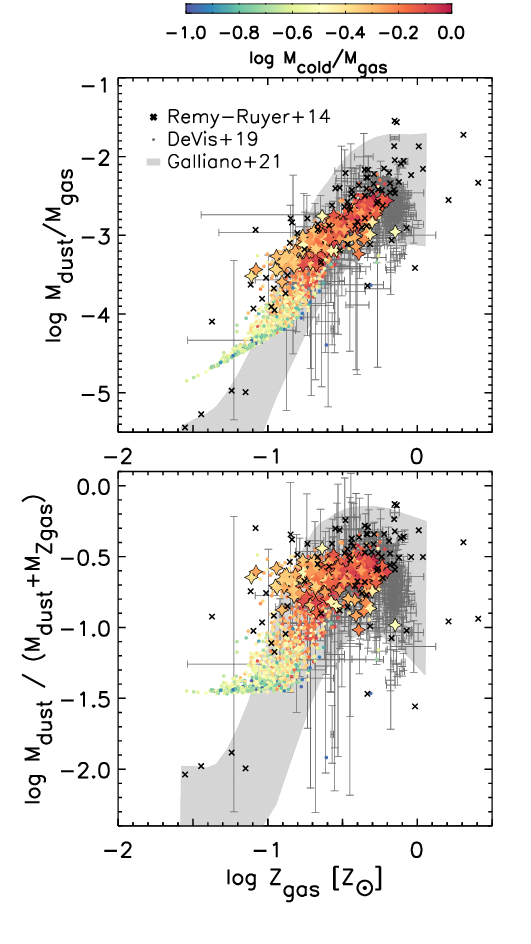}
\caption[width=\textwidth]{
Dust-to-Gas (\textit{top panel}) and Dust-to-Metals (\textit{bottom panel}) ratio as a function of the ISM gas phase metallicity $Z$ for simulated galaxies at $z=0$ (MWHM galaxies are star symbols). The compilations of observational data are by \cite{Remy-Ruyer2014} (black cross) and \cite{Vis2019} (dots). We also report the fit derived by \cite{Galliano2021a} (grey shaded area, which contains the 98$\%$ of their observations). The color code shows the cold fraction of the total gas content of each galaxy.
}
\label{fig:mdZ}
\end{figure}

\begin{figure}
\centering
\includegraphics[width=0.4\textwidth]{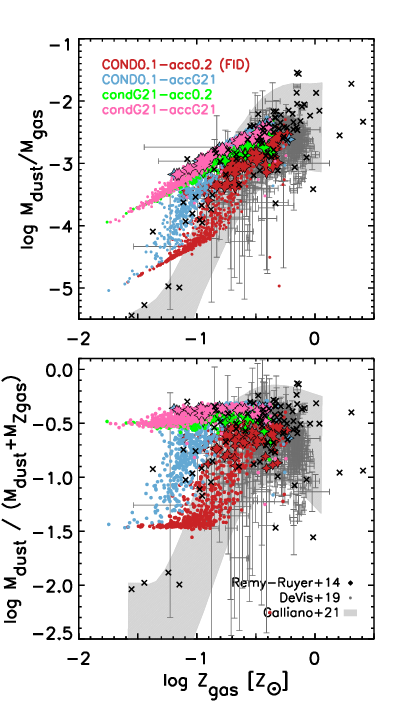}
\caption[width=\textwidth]{Same as Fig. \ref{fig:mdZ}, but for different assumptions on the dust condensation and/or accretion efficiencies (see Section \ref{sec:dtg})}
\label{fig:mdZ_models}
\end{figure}

\subsubsection{Dust and Metallicity}
\label{sec:dtg}
While both the gas phase and dust metals originate from evolved stars, their relative abundances are strongly affected by dust evolutionary processes in the ISM. We show in Figure \ref{fig:mdZ}, both the dust-to-metals (DTM; bottom panel) and dust-to-gas (DTG; top panel) ratio as a function of the metallicity at $z=0$.
Simulated galaxies distribute in these diagrams following three distinct regimes, previously pointed out by several works \cite[e.g.][and references therein]{Hirashita2013}. A low metallicity one in which the direct production from stars determines the dust content. An intermediate metallicity regime characterized by a much more rapid growth of the \revdue{DTG} ratio with the metallicity due to accretion. A high metallicity regime, sometimes referred to as {\it saturation limit}, in which the growth of the \revdue{DTG} ratio slows down again, because the depletion of metals that can contribute to the dust budget is almost complete. The role of grains accretion in shaping this relation is highlighted by the fraction of cold gas in each galaxy, color coded in Fig. \ref{fig:mdZ}. Grains growth occurs in the cold and dense phase of the ISM, indeed galaxies with higher cold gas fraction at a given metallicity tend to have a larger DTG and DTM.

A few model assumptions largely determine the exact locations of the three aforementioned regimes. This dependence can be appreciated by inspecting Figure \ref{fig:mdZ_models}, wherein we show, besides our present fiducial model, results obtained adopting different dust parameter values, taken from \cite{Granato2021}, and summarized in Table \ref{tab:DustModels}.
The vertical position of the low metallicity branch depends on the adopted dust condensation efficiencies for the evolved stars ejecta.
The transition between the low and intermediate metallicity ones depends on the adopted accretion efficiency, while the vertical position of the saturation limit is dictated by the maximum metal availability, in turn  related to the adopted IMF and stellar yields. To get a good match with available data of nearby galaxies, we had to reduce substantially both efficiencies with respect to those assumed by \cite{Granato2021}. However, we reiterate that these changes do not affect significantly the results discussed in the latter paper (see Figure \ref{fig:gra21}).
Figure \ref{fig:mdZ} displays a fairly good agreement between the fiducial model and the available data.  However, the three regimes are much less evident in the data, possibly due to their significant uncertainties.
Before concluding this section, we point out that if we adopt the alternative MUPPI model effDTG proposed in Section \ref{sec:SFRD}, in which galaxies turn out to be a lower amount of gas in which accretion can efficiently occur, we need a higher accretion efficiency, closer to that adopted in \cite{Granato2021} to match this dataset \rev{(see Appendix \ref{app:effdsg})}.

\subsubsection{Size and chemical composition of grains}
\label{sec:scaling_size_comp}
\begin{figure*}
\centering
\includegraphics[width=0.9\textwidth]{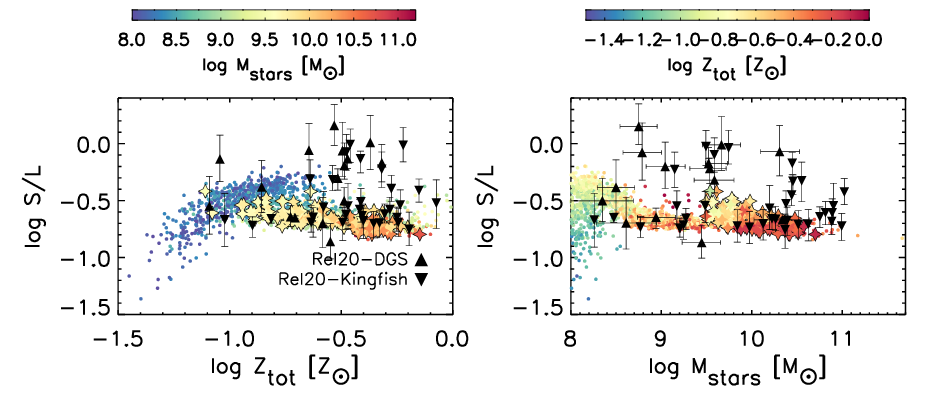}
\caption[width=\textwidth]{Small-to-Large grains ratio at $z=0.0$ as a function of total ISM metallicity $Z_\text{tot}$ and stellar mass. Star symbols represent MWHM galaxies. \rev{Local observations from \cite{Relano2020} are shown for comparison (upward triangles for the DSG sample, downward triangles for Kingfish galaxies)}.}
\label{fig:Small_to_Large}
\end{figure*}

\begin{figure*}
\centering
\includegraphics[width=0.9\textwidth]{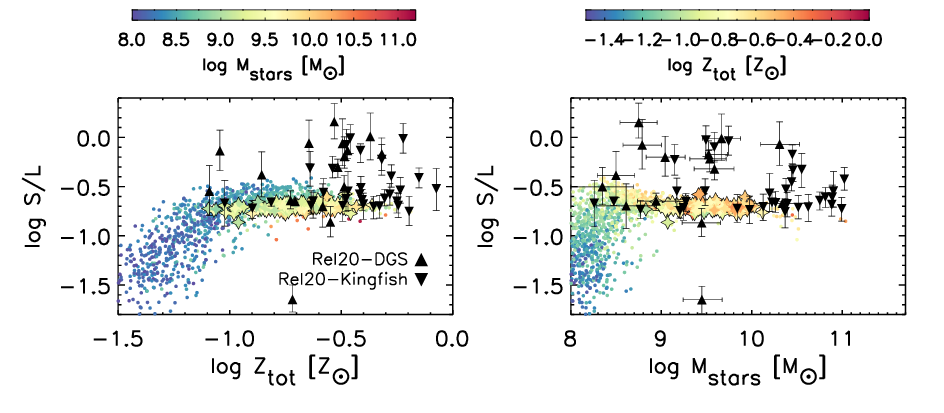}
\caption[width=\textwidth]{Same as Fig. \ref{fig:Small_to_Large} at $z=1.25$. Star symbols refer to MWHM progenitors. \rev{Local data from \cite{Relano2020} are also shown for reference.}}
\label{fig:Small_to_Largez1}
\end{figure*}

\begin{figure*}
\centering
\includegraphics[width=0.9\textwidth]{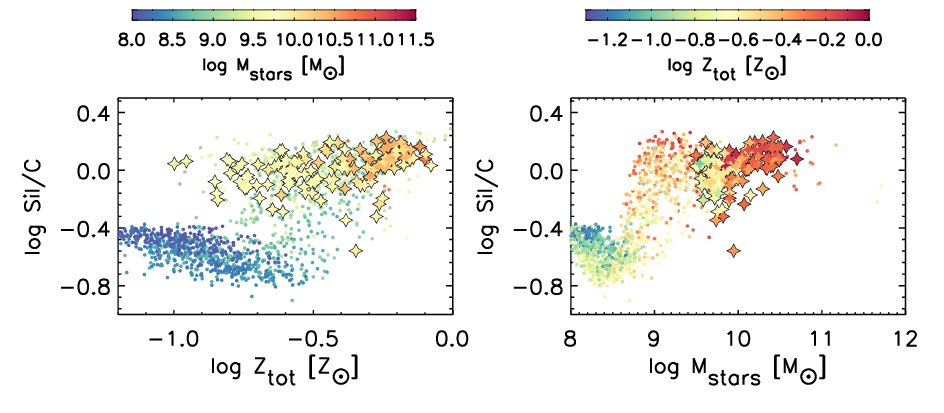}
\caption[width=\textwidth]{Same as Figure \ref{fig:Small_to_Large}, but the Silicate-to-Carbon grains ratio is shown.}
\label{fig:Sil_to_Car}
\end{figure*}

\begin{figure*}
\centering
\includegraphics[width=0.9\textwidth]{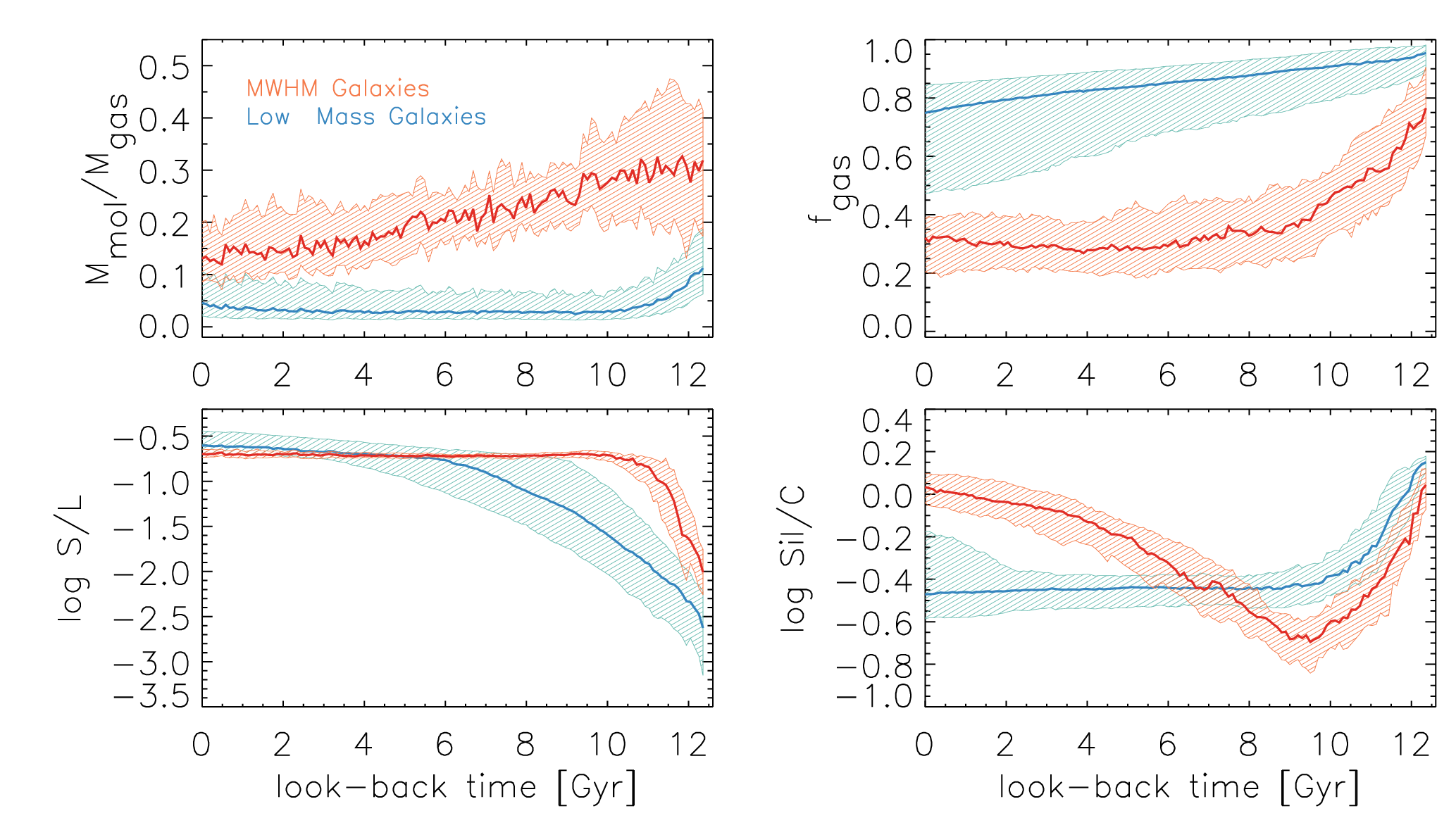}
\caption[width=\textwidth]{Evolution in time of the median and 25-75\% percentiles of molecular fraction, gas fraction, Small-to-Large grains ratio and Silicate-to-Carbon grains ratio.
MWHM galaxies are depicted in red while in blue we show the behaviour of galaxies with $M_{200} < 3\times 10^{11} M_{\odot}$.}
\label{fig:low_high_evo}
\end{figure*}

The dust model follows the formation and evolution of four different grain species, tracking the different chemical composition and, to a minimal level, size distribution of astrophysical dust. The two panels of Figure \ref{fig:Small_to_Large} show the Small-to-Large grains mass ratio \rev{(S-to-L)} respectively as a function of the {\it total} ISM metallicity $Z_\text{tot}$ (that is including both metals in gas and in dust grains) and the stellar mass \rev{at z=0}. As in the previous plots, large symbols refer to MWHM galaxies. This Figure can be directly compared to figure 6 of \cite{Hou2019}, who also performed cosmological simulations in a (larger) cosmological volume, following dust enrichment by means of the two-size approximation, but without treating dust chemical composition. Our results broadly agree with those presented in the latter paper. The major differences are because our local galaxy sample covers a significantly narrower $Z_\text{tot}$ interval. On the high $Z_\text{tot}$ (or $M_\text{stars}$) end, this could be related to the fact that they adopted the significantly more top-heavy IMF by \cite{Chabrier2003}. At the low $M_\text{stars}$ and $Z_\text{tot}$ end, some more subtle differences in their modelling resulted in a significantly lower metallicity, as it can be appreciated by comparing their Figure 2 with our Figure \ref{fig:Zgas_Ms_M200}. We point out that our results in this mass range are in better agreement with observations. The behaviour of our model in this very low metallicity regime can be explored at earlier time, as shown in Figure \ref{fig:Small_to_Largez1} for redshift 1.25. Taken together, these two figures confirm the interpretation put forward by \cite{Hou2019}. As long as galaxies are small, with high gas fraction and low metallicity, the relatively small dust content is dominated by large grains produced by stellar ejecta. Increasing the metallicity, and as such the mass, small grains production by shattering and accretion becomes more and more important, and the \revdue{S-to-L} grains mass ratio increases quickly with $Z_\text{tot}$. A further increase of metallicity above $\text{log} \, Z_\text{tot} \sim -1.$, or $\text{log} \, Z_\text{gas} \sim -1.5$, does not produce an increase of S-to-L, because in the saturation regime the mass increase of small grains due to accretion turns out to be balanced by their coagulation into large ones.
\rev{We compare our predictions with the derivation of the \revdue{S-to-L} grains mass ratio presented in \cite{Relano2020}, where the authors exploited the integrated SEDs of the KINGFISH and DSG surveys. Our results as a function of stellar mass and metallicity are in reasonable agreement with data, but for the lack of simulated galaxies with log S-to-L $\gtrsim -0.2$. Moreover, observations seem to confirm the predicted flat S-to-L vs $M_\text{stars}$ relationship up to large stellar masses. By converse, \cite{Hou2019} predicted a step decrease of the ratio at high stellar mass, tentatively ascribed by \cite{Relano2020} to the form of AGN feedback in the latter simulation. (see discussion in Sec. 4 of \citealt{Relano2020}). \\}
Similarly, Figure \ref{fig:Sil_to_Car} shows the dependence of the \revdue{Silicate-to-Carbon} dust mass ratio \rev{(Sil-to-C)} at $z=0$ on total metallicity and stellar mass. Interestingly, in the $\text{log} \, Z_\text{tot}$ interval $[-0.9,-0.6]$  galaxies tend to have a broad range of Sil-to-C, with a pretty dichotomous distribution. Less massive galaxies with $M_\text{stars} \lesssim 3 \times 10^{8} \rm{M}_\odot$ feature a ratio about a factor 3 smaller than MWHM galaxies at the same $Z_\text{tot}$, and similar to that characterizing less metal-rich ones, in their same stellar mass range.
The different behaviour in the two stellar mass regimes can be better understood by inspecting Figure \ref{fig:low_high_evo}, wherein we show, for two relevant mass ranges, the evolution of the gas molecular fraction $M_\text{mol}/M_\text{gas}$ (as given by the MUPPI model), the gas mass fraction $f_{gas}=M_\text{gas}/(M_\text{gas}+M_\text{stars})$, the \revdue{S-to-L} and \revdue{Sil-to-C} dust mass ratio.
At very early time, all galaxies in our simulation tend to have a relatively large  Si-to-C ratio, because dust is dominated by grains produced by SNII. After that, the ratio decreases quickly due to the copious production of C dust by AGBs. If accretion begins to be significant, the ratio starts to increase again, but slowly because silicate dust growth is controlled by the limited availability of the key-element(s) \citep[see also section 3 in][]{Granato2021}. Lower mass galaxies have more difficulty reaching the accretion dominated regime than more massive ones at the same metallicity because they have a smaller fraction of dense gas, in which accretion occurs.  In the low mass bin, the gas mass fraction is always higher by a factor of a few, but the fraction of molecular gas turns out to be lower by an order of magnitude or more during most of cosmic history (top panels of Figure \ref{fig:low_high_evo}). As a result, the dust accretion process never becomes strong enough to increase the Sil/C ratio in low mass galaxies.

Finally, we remark that the typical values obtained here for both the Large-to-Small and Sil-to-C ratios for MWHM galaxies at $z=0$ are not far from those adopted in most dust mixture models derived for the MW dust. For instance, model 4 in \cite{Weingartner2001} has S-to-L $\simeq 0.21$ and Sil-to-C $\simeq 2.5$.  However, the variations predicted in different galaxies at $z=0$, and even more at higher redshift, would be significant when computing dust radiative effects \citep[see also section 3.1 in][]{Granato2021}.

%
%

\subsection{Dust outside galaxies}
\label{sec:dustoutgalaxies}
Observations reveal the presence of dust outside the {\it galactic region} (\citealt{Holwerda2009}, \citealt{Menard2010}, \citealt{Roussel2010}, \citealt{Hodges-Kluck2014}, \citealt{Peek2015}, \citealt{Ruoyi2020}), and this presence has also been found in a few cosmological simulations \citep{McKinnon2017,Aoyama2018}. Exploring this issue is relevant to describe more completely the dust and metals cosmic cycle, to constrain the cooling capability of the CGM, and to characterize the opacity of the Universe on intergalactic scales. The dust produced and re-processed in galactic environments may indeed be ejected out of the ISM thanks to various mechanisms, such as galactic winds and radiative pressure \citep[see e.g.][]{Bianchi2005,Hirashita2019a}. Among these processes, only galactic winds acting on gas particles are taken into account in our simulation. Moreover, in our framework dust is dynamically coupled with gas. This assumption, which could affect the amount and nature of ejected dust, has been found to be a good approximation in most cases, as a result of the strong coupling between these two components in most situations \citep[see e.g.][who explicitly treat dynamical forces acting on dust particles]{McKinnon2018}.\\
In the following sections we analyze our simulation to investigate the amount and the properties of dust around our simulated galaxies at different epochs.

\subsubsection{Dust profile around galaxies}
\label{sec:profile}

To quantify the amount of dust beyond galactic scales we compute the dust surface density profiles. From the observational point of view, \cite{Menard2010} addressed this problem by studying the reddening due to dust associated with the halo (i.e. on extra-galactic scales) of massive galaxies. They performed this study by exploiting the properties of a sample of $\sim 85000$ quasars at $z \gtrsim 1$ and a huge sample of $2 \cdot 10^7$ galaxies at $z \simeq 0.3$. They cross-correlated the colours of the quasars with the (projected) overdensity of galaxies, constraining in this way the reddening due to grains outside galactic regions. Their conclusion is that the dust associated with an extra-galactic diffuse component is at least comparable with that observed in the ISM of galaxies, and that the dust profile extends up to $\simeq 10 \, \text{Mpc}$ from the centre of galaxies. Later studies \citep[e.g.][]{Peek2015} confirmed the presence of a large amount of dust beyond galactic radii.

\begin{figure}
\centering
\includegraphics[width=1.15\columnwidth]{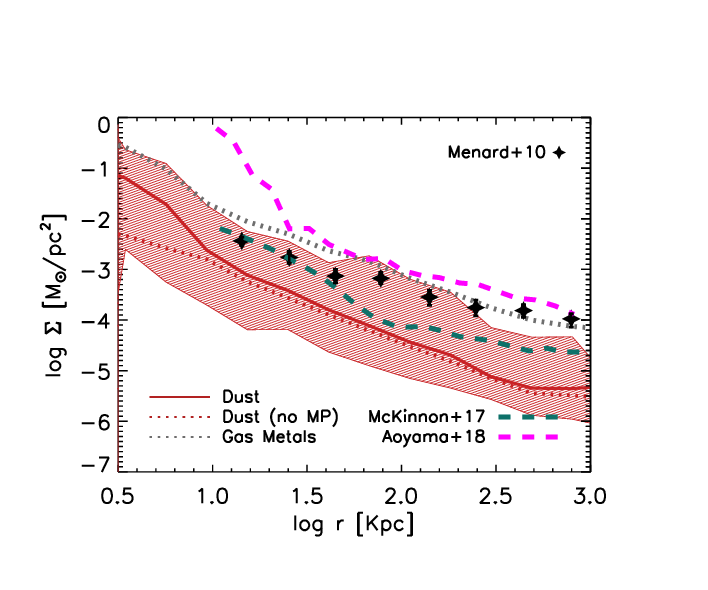}
\caption[width=\textwidth]{Median dust surface density profile (solid line) around galaxies with $M_\text{stars}>10^{10} \, M_\odot$ at $z=0.3$, with the shaded area indicating the full dispersion of the profiles distribution. The profile excluding MultiPhase gas particles is also shown as a red dotted line, as well as the gas metals profile (grey dotted line). Filled star points data refer to observations by \cite{Menard2010}. We also overplot results from \cite{McKinnon2017} (green dashed line) and \cite{Aoyama2018} (magenta dashed line).}
\label{fig:profile}
\end{figure}

To compare the results of our simulation with their observational estimates, we compute the dust surface density considering galaxies at $z \simeq 0.3$ with $M_\text{stars} > 10^{10} \, M_\odot$; this range is broadly similar to the masses of the galaxies studied by \cite{Menard2010} \citep[see e.g.][]{Aoyama2018}. We note that this mass range is actually a bit larger than the one featured by our MWHM galaxies progenitors at $z=0.3$. However, using MWHM progenitors as in the rest of the paper, would only slightly modify the obtained profile, without affecting our conclusions. We present results in Figure \ref{fig:profile}, where we show the median dust profile\footnote{Radial profiles have been obtained considering a randomly oriented cylinder centered on each of the selected galaxies. We group gas particles in $20$ logarithmic bins along the cylinder radius, and we compute the surface density by summing the contribution of all the particles in each annulus. The vertical axis coincides with the $z$ axis of our simulated box. The radius of the cylindrical region is $4$ cMpc$/$h, while its half-height $500$ ckpc$/$h; we verified that our results do not critically depend on the choice of these parameters.} compared with  \cite{Menard2010}. We also overplot the median dust profile obtained excluding MultiPhase particles, as well as the gas metals profile.
The slope of our dust profile is a bit steeper than the one derived by \cite{Menard2010}, while its normalization is slightly lower than the data, although this tension is reduced when considering the whole, relatively large, dispersion.
We point out that the dust density derived from reddening by \cite{Menard2010} depends also on the dust model adopted: while they assume Small Magellanic Cloud dust type, actually the properties of dust on extragalactic scales at $z=0.3$ may differ from the ones inferred in local galaxies (see the following discussion). \\
To understand to what extent this profile originates from the diffuse dust component outside galaxies, we plot in Fig. \ref{fig:profile} also the median dust profile obtained excluding
multiphase particles, which trace the dusty galactic environments in MUPPI. The contribution of non-multiphase particles to the total profile is minor within $r \lesssim 10 \, \text{kpc}$ (which is the galactic region where \rev{star formation} occurs), while outside this region it is about the 90\% of the total. Therefore, the dust budget in the simulations on extragalactic scales is dominated by a diffuse component of dust in galactic halos. Satellites and overlapping galaxies play a minor role, in agreement with the conclusions by \cite{Menard2010}.
Finally, we compare our profile with results from other two cosmological simulations, including treatment of dust evolution by \cite{McKinnon2017} and \cite{Aoyama2018}. Noticeably, despite the different prescriptions adopted in the sub-grid modelling of dust evolution, all the three simulations show a similar slope of $\Sigma_\text{dust}$ on extra-galactic scales. In particular, our profile normalization is quite close to that in \cite{McKinnon2017}. We remark that not only dust-related prescriptions may affect the profile. Feedback and galactic winds recipes, among the others, are expected to have a substantial impact on these results.

\subsubsection{Redshift evolution of dust profiles}

\begin{figure*}
\centering
\includegraphics[width=0.8\textwidth]{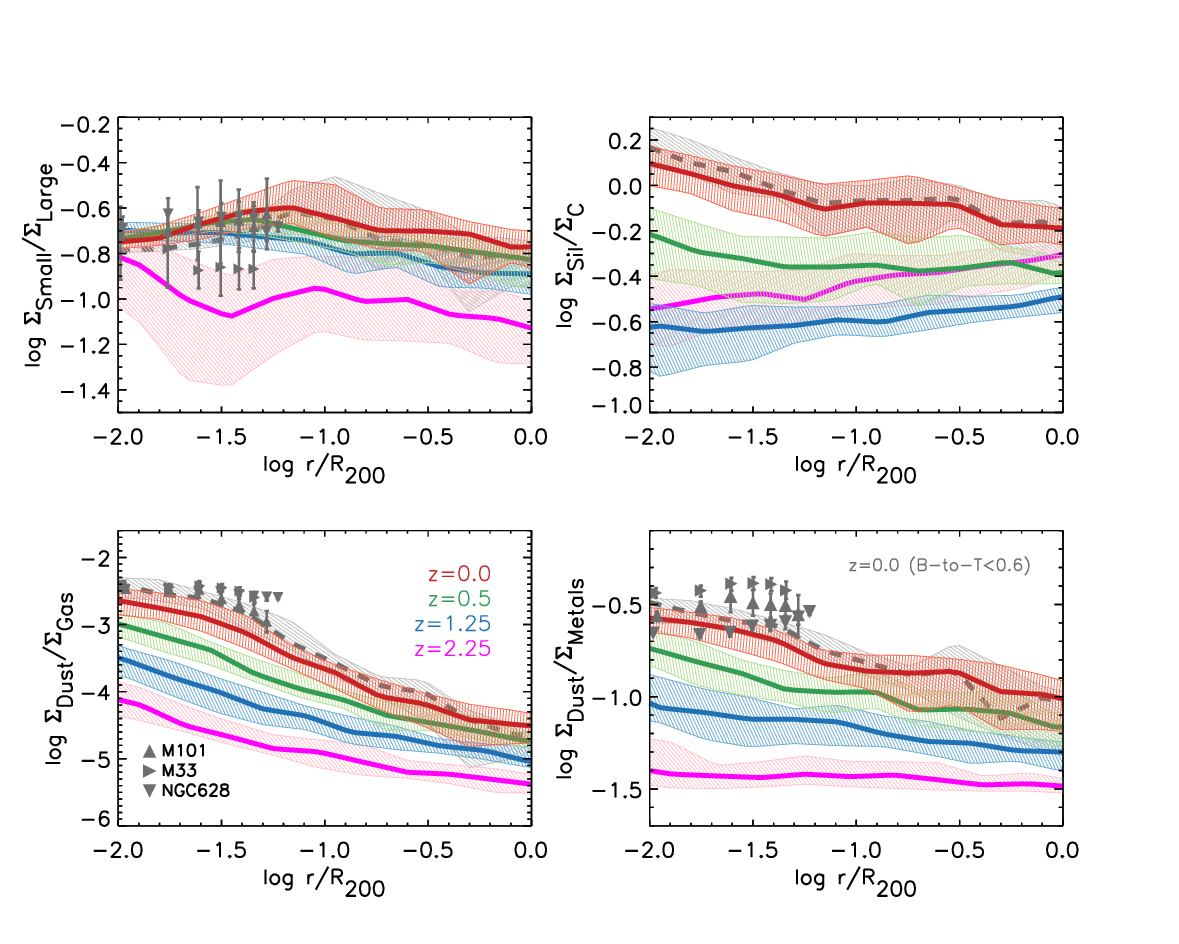}
\caption[width=\textwidth]{Median radial profiles ratio of Small-to-Large (top left), Silicate-to-Carbon (top right), Dust-to-Gas (bottom left) and Dust-to-Metals (bottom right) for MWHM galaxies ($z=0.0$) and their progenitors ($z=0.5$, $z=1.25$, and $z=2.0$). The radial coordinate has been normalized to the $R_{200}$ of each galaxy. We show the 25-75th percentile dispersions. \rev{We over plot the profiles reported by \cite{Relano2020} derived from the spatially resolved observations of three local disc galaxies (M101, M33, NGC628). For the sake \revdue{of} a more meaningful comparison, we also show the profiles obtained considering only $z=0.0$ simulated disc galaxies (B-to-T$< 0.6$, grey line and shaded region).}}
\label{fig:profile_zevo}
\end{figure*}

In Fig. \ref{fig:profile_zevo} we show the redshift evolution of the median profiles of the \revdue{S-to-L} and \revdue{Sil-to-C} dust mass ratios, as well as those of DTG and DTM.
In this analysis we consider the MWHM galaxies and their main progenitors at $z=0.5, 1.0,$ and $2.0$. The radial coordinate has been normalized to the $R_{200}$ of each halo at the corresponding redshift. Both the DTG and DTM show a profile that increases monotonically with time, as a result of the continuous production and ejection of dust from galaxies.

\begin{figure}
\centering
\includegraphics[width=0.95\columnwidth]{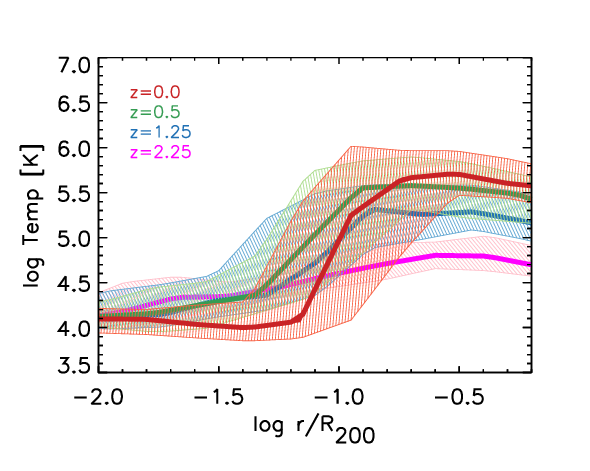}
\caption[width=\textwidth]{Median temperature profile of MWHM galaxies and their progenitors at $z=0.5$, $z=1.25$, and $z=2.25$. Shaded regions are the 25th-75th percentiles dispersion.}
\label{fig:profile_temp}
\end{figure}

Large grains dominate both inside and outside ($r>0.1\,R_{200}$) galaxies at any redshift. At $z \gtrsim 2$ the \revdue{S-to-L} grains ratio is significantly smaller outside, and this difference decreases with time, almost vanishing at $z=0$. This is related to the rapid increase of this ratio in the galactic region at early time (look back time $\gtrsim 10$ Gyr) for massive galaxies (see red band in lower left panel of Figure \ref{fig:low_high_evo}). Dust is primarily produced in the galactic region and transported outside. It takes some time to pollute the {\it outside} region. Therefore dust properties in latter regions reflects the properties of dust in the former one at somewhat earlier time.
At low redshift, the ratio tends to increase up to $\log r/R_{200} \simeq -1.2$. This is likely due both to the radial decrease of the specific star formation, so that the relative importance of small grains production by accretion with respect to that of large grains by stellar injection increases, and to a progressive shift of the balance between shattering and coagulation in favour of shattering, again related to a fractional decrease of the dense, star-forming and dust coagulating ISM.
At larger radii, the ratio decreases because the gas suddenly reaches a typical temperature of a few $10^5$ K, at which sputtering, whose efficiency is ten times larger for small grains, begins to be non-negligible, progressively with time (Figure \ref{fig:profile_temp}).

As for the contribution of Silicates and C to the dust budget, Fig. \ref{fig:profile_zevo} (top right panel) confirms the trend already discussed in Sec. \ref{sec:scaling_size_comp}. The importance of the former kind of grain increases with cosmic time at $z \lesssim 2$ or look back time $\lesssim 10$ Gyr in massive enough galaxies. This evolution occurs both in the galactic region (see also Fig \ref{fig:low_high_evo} bottom right panel) as well as outside it. It arises from the delay imposed to silicate accretion by the limited availability of the key element at early times. A striking feature of the profile evolution is the reversal of the slope with time. It switches from increasing Sil/C with radius at high-z to low-z, where the opposite occurs. Again, this can be understood taking into account that the dust mixture outside the galaxies is imprinted, with some reprocessing, by the conditions of the dust inside the galaxies at some earlier time. Fig \ref{fig:low_high_evo} demonstrates that Sil/C rapidly decreases with time at $z \gtrsim 2$ (look-back time $\gtrsim 10$ Gyr) down to a minimum, and then in massive galaxies, it begins to increase slowly.

\rev{We compare our results with the a few available observations at $z=0.0$. \cite{Relano2020} used spatially resolved observations of three local disc galaxies (M101, M33, and NGC628) to study the shape of the S-to-L, DTG and DTM radial profiles\footnote{They express the radius in units of $R_{25}$, the radius where the B-band surface brightness falls to $25 \, \rm{mag} \rm{arcsec}^{-2}$. To translate their profiles to our $R_{200}$ normalization of the radius, we first assumed $R_{25}=4R_{d}$, where $R_{d}$ is the disc-scale length obtained by fitting the simulated stellar surface density profile with an exponential function $\Sigma(r) = \Sigma_0 \rm exp (-r/R_d)$. Then we used the mean $R_{25}/R_{200}$ for our MWHM disc-like galaxies (those with B-to-T $<0.6$) at $z=0.0$, which turns out to be $\simeq 0.07$.  }.
Since in the latter work only disc galaxies have been considered, we show at $z=0.0$ also the profile of simulated MWHM galaxies with B-to-T$<0.6$ (grey lines and shaded areas).  The simulation reproduces reasonably well the observed profiles. The DTG and DTM profiles are underestimated by a modest factor $\simeq 1.6$, but the discrepancy reduces for the B-to-T $<0.6$ sample. Two out of the three observed S-to-L profiles have large uncertainties, and show a quite constant profile with log S-to-L $\simeq -0.7$, as well as our simulated $z=0.0$ galaxies. The S-to-L profile of M33 is a factor $\simeq 1.6$ below our results.}

\section{Summary and Discussion}
\label{sec:conclusions}
We present and discuss hydrodynamical simulations performed with the dust evolution model proposed by \cite{Granato2021}. The latter implements the two size approximation to predict the formation and evolution of dust, taking into account its chemical composition and, to a minimal level, the size distribution. The production of grains by stellar sources and the subsequent evolution of grains in the galactic ISM via different processes are considered. As for the star formation and feedback modelling, we adopt the MUPPI formalism, here in the specific version presented by \cite{Valentini2020}, without any tuning of its parameters.
Nevertheless, we propose a couple of possible directions to relieve the tensions with some properties of the galaxy population. While \cite{Granato2021} discussed zoom-in simulations expected to produce Milky Way type galaxies, here we consider cosmological boxes to test and adjust the dust model on a relatively large sample of galaxies.
We mainly focused on the dust properties of Milky Way-like galaxies since MUPPI has been calibrated to reproduce the properties of these objects. More precisely, we devoted particular attention to galaxies formed in dark matter halos having a mass $M_\text{200}$ within a factor 3 that of estimated for the MW, dubbed Milky Way Halo Mass galaxies (MWHM).
Dust-independent properties, such as the morphology, the gas fraction and the metal content of our sample of simulated MWHM galaxies, are in acceptable agreement with local observations. However, we found clear tensions between our results and data when considering the whole simulated population of galaxies. The predicted cosmic \revdue{SFRD} is a factor of $\sim 3$ lower than the observed one at $z \sim 2-3$. We propose a possible modification of the star formation prescriptions to improve this issue: a star formation efficiency that depends on the \revdue{DTG} ratio (Eq. \ref{fstarDSG}). Such a solution produces a SFRD in significantly better agreement with the data.
We also pointed out a dearth of quenched galaxies in the local Universe, a problem that can be cured by adopting a more effective SMBH growth and feedback. However, the introduction of these modifications would require a re-tuning of the MUPPI parameters. For this reason we prefer to keep the setup carefully calibrated by \cite{Valentini2020} on zoom-in simulation and leave for the future a complete analysis of the issue.

Instead, we do change the dust parameters adopted in \cite{Granato2021}, namely the condensation efficiency in stellar ejecta and accretion timescale. Such a modification does not significantly affect MWHM galaxies at $z \lesssim 1$, but it is needed to reproduce the locally observed relation between the dust content and metallicity (DTG-Z and DTM-Z, Fig. \ref{fig:mdZ}). These relations are crucial to understand the relative contribution of stellar production of dust and grains accretion in the ISM. In particular, the accretion timescale mainly drives the \textit{transition metallicity} where accretion becomes dominant. At the same time, a low condensation efficiency allows for matching the DTG ratio of the lowest metallicities objects. After the tuning, all the simulated scaling relations involving dust, stars and metals content appear consistent with the available observations.
Our dust model also predicts the relative abundance of small and large grains and their chemical composition (silicates and carbonaceous). Interestingly, these properties are partly entangled. Low mass galaxies, where stellar production drives the dust content, are dominated by large, carbonaceous grains. Silicates, which in our model follow an olivine-like chemical composition, dominate the mixture when accretion on small grains becomes relevant (galaxies with $M_* \gtrsim 10^{8.5} \, M_\odot$, $\, \text{log}Z/Z_\odot \gtrsim -0.8$). Moreover, in this transition regime, the Small/Large ratio reaches a maximum, which is suppressed in larger galaxies because of the intervening contribution of coagulation. MWHM galaxies appear to be silicates and large grains rich at $z=0.0$. Dust evolution in these objects shows that silicates growth becomes relevant at $z \lesssim 2$, at variance with low mass galaxies, which have an order of magnitude lower molecular gas mass fraction (which determines the accretion efficiency) for the whole cosmic time. For MWHM galaxies, the final S/L and Sil/C values are broadly consistent with those of the most common mixtures adopted in dust extinction models. Nevertheless, we remark here that a deeper analysis, e.g. the study of depletion patterns, would be needed to test our chemical predictions, which are in turn strongly influenced by our assumptions on silicates composition, as well as by the adopted IMF.\\

Surface density profiles around massive galaxies agree reasonably well with the few available observations, confirming the clues from literature about a non-negligible dust abundance beyond galactic regions. Moreover, the redshift evolution of the S/L and Sil/C profiles shows that halo dust reflects the properties of galactic dust at earlier times, an effect that is more evident at high redshift. \\

The issues discussed here suggest future developments. We need to investigate the weaknesses of our sub-resolution model MUPPI, eventually introducing new prescriptions and re-tuning its parameters to obtain reasonable results for the whole galaxies population across cosmic time. This study requires proper care to keep the good performances of MUPPI when adopted in zoom-in simulations of disk galaxies. Moreover, we remark that we adjusted the dust parameters on local observations. Once we have a more satisfactory galaxy evolution framework, it would be interesting to assess the capability of our model to correctly reproduce dust properties at high redshift.


\section*{Acknowledgements}
We are indebted to an anonymous referee for careful reading and for comments which have helped improve the quality of this work.
We thank Pieter De Vis and Mónica Relaño for providing us with data used for comparison to our model predictions.
We also warmly thank Pierluigi Monaco and Laura Silva for useful discussions.
This project has received funding from the Consejo Nacional de Investigaciones Cient\'ificas y T\'ecnicas de la Rep\'ublica Argentina (CONICET) and from
the European Union's Horizon 2020 Research and Innovation Programme under the Marie Sklodowska-Curie grant agreement No 734374.
Simulations have been carried out at the computing centre of INAF (Italia). We acknowledge the computing centre of INAF-Osservatorio Astronomico di Trieste, under the coordination of the CHIPP project \citep{bertocco2019,Taffoni2020}, for the availability of computing resources and support.
MV is supported by the Alexander von Humboldt Stiftung and the Carl Friedrich von Siemens Stiftung. MV also acknowledges support from the Excellence Cluster ORIGINS, which is funded by the Deutsche Forschungsgemeinschaft (DFG, German Research Foundation) under Germany's Excellence Strategy - EXC-2094 - 390783311.
SB acknowledges financial support from Progetti di Rilevante Interesse Nazionale  (PRIN) funded by the Minisitero della Istruzione, della Università e della Ricerca (MIUR) 2015W7KAWC, the Istituto Nazionale di Fisica Nucleare (INFN) InDark grant. AB and AL acknowledge funding from the PRIN MIUR 2017 prot. 20173ML3WW, `Opening the ALMA window on the cosmic evolution of gas, stars and supermassive black holes'.

\section*{Data Availability}
The data used for this article will be shared on reasonable request to the corresponding author.



\bibliographystyle{mnras}




\FloatBarrier
\appendix

\section{Runs with enhanced BH feedback}

\label{app:morebh}

\begin{figure}
    \centering
    \includegraphics[width=0.9\columnwidth]{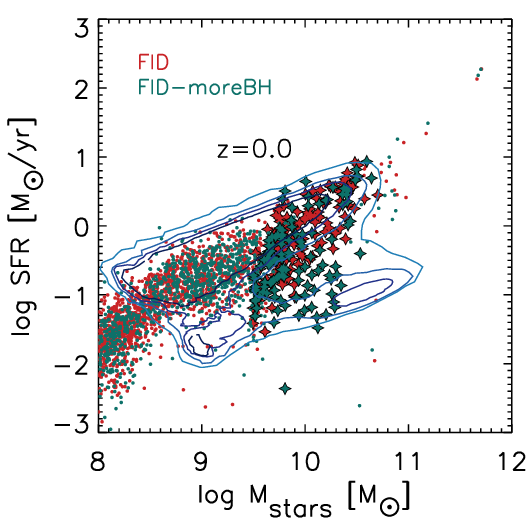}
    \caption{Same as left panel of Fig. \ref{fig:mainsequence} for our fiducial run and the run with enhanced BH feedback (see text). MWHM galaxies are marked with star symbols.}
    \label{fig:mainsequence-moreBH}
\end{figure}

\begin{figure}
\centering
\includegraphics[width=0.9\columnwidth]{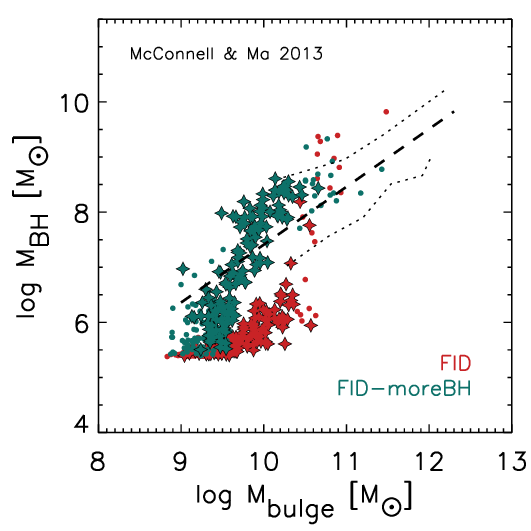}
\caption[width=\textwidth]{$M_\text{bulge}-M_\text{BH}$ at $z=0.0$ for fiducial run and the run with enhanced BH feedback. MWHM galaxies are marked with star symbols. The fit by \cite{McCon13} is shown.}
\label{fig:BHMbulge}
\end{figure}

The model dubbed \textit{FID-moreBH} was obtained from the fiducial one by increasing the feedback efficiency  $\epsilon_\text{f}$, from the value 0.01 used by \cite{Valentini2020} (see their Eq. 19 and 25, and their Table 2) to 0.05, and by switching off the angular momentum limiter described in their section 3.3. This modification produces a population of more quenched galaxies at low redshift \revdue{(Figure \ref{fig:mainsequence-moreBH})}, and yields a correlation between the SMBH and spheroidal stellar component in better agreement with the observations \revdue{(Figure \ref{fig:BHMbulge})}, with negligible modifications of the other properties discussed in the body of the paper.

\section{Runs with DTG dependent Star Formation}

\begin{figure*}
\centering
\includegraphics[width=1.\textwidth]{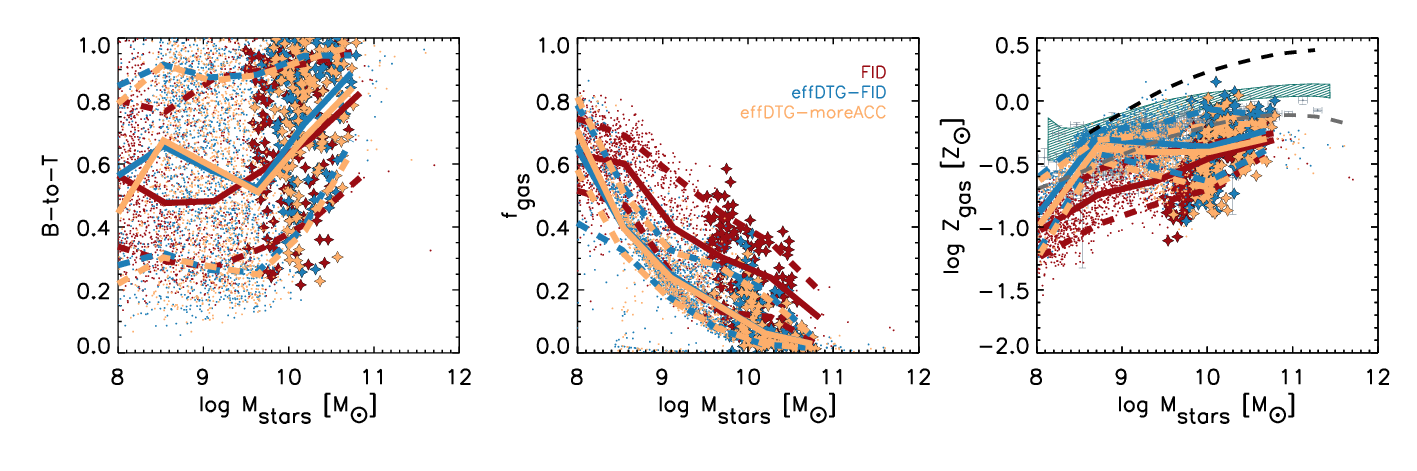}
\caption[width=\textwidth]{Bulge-to-Total ratio (left panel), gas fraction (middle panel) and gas phase metallicity (right panel) as a function of the stellar mass for the fiducial model (red) and two different models where we introduced the star formation efficiency dependent on DTG (Section \ref{sec:SFRD}): \textit{effDTG-FID} (blue) and \textit{effDTG-moreACC} (yellow). See the text for details. The solid line marks the median of all galaxies, while the \revdue{dashed lines enclose the 25-75th percentiles}. Star symbols refer to MWHM galaxies. \revdue{Observational data in the right panel are the same as in Fig. \ref{fig:Zgas_Ms_M200}.}}
\label{fig:BT_fg_Z_effDTG}
\end{figure*}


\begin{figure*}
    \centering
    \begin{subfigure}[t]{0.3\textwidth}
        \centering
        \includegraphics[width=\linewidth]{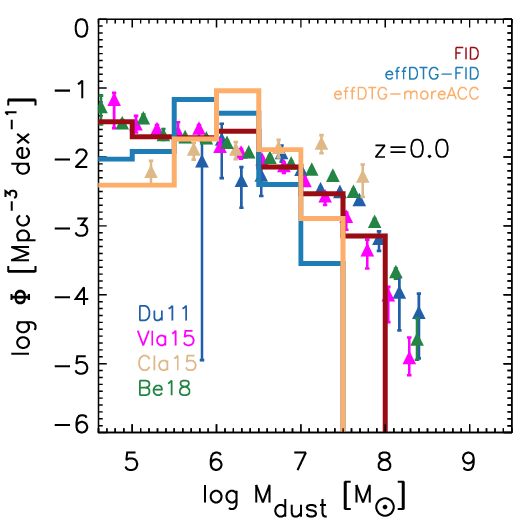}
    \end{subfigure}
    \hfill
    \begin{subfigure}[t]{0.3\textwidth}
        \centering
        \includegraphics[width=\linewidth]{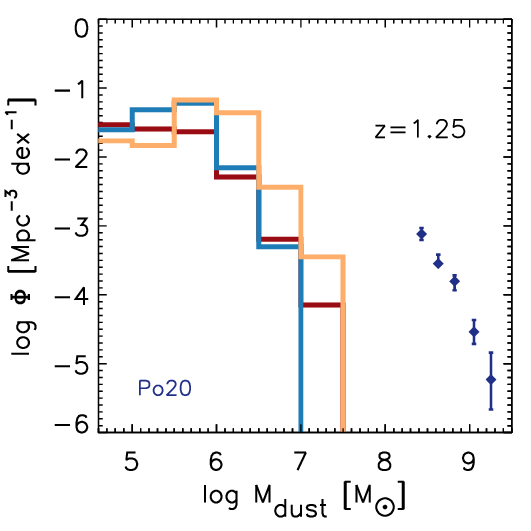}
    \end{subfigure}
    \hfill
    \begin{subfigure}[t]{0.3\textwidth}
        \centering
        \includegraphics[width=\linewidth]{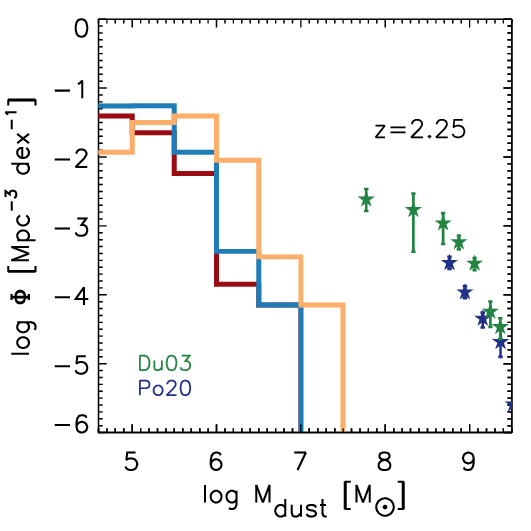}
    \end{subfigure}

    \caption{Dust Mass Function of the fiducial run (red) compared to the \textit{effDTG-FID} (blue) and \textit{effDTG-moreACC} (yellow) models at different redshifts. We show observations from \citet{Dunne2011}, \citet{Vlahakis2005}, \citet{Clark2015}, \citet{Beeston2018} ($z=0$), \citet{Pozzi2020} ($z=1.25$ and $z=2.25$) and \citet{Dunne2003} ($z=2.25$).}
    \label{fig:dustMF_effDTG}
\end{figure*}

\begin{figure*}
    \centering
    \includegraphics[width=1.5\columnwidth]{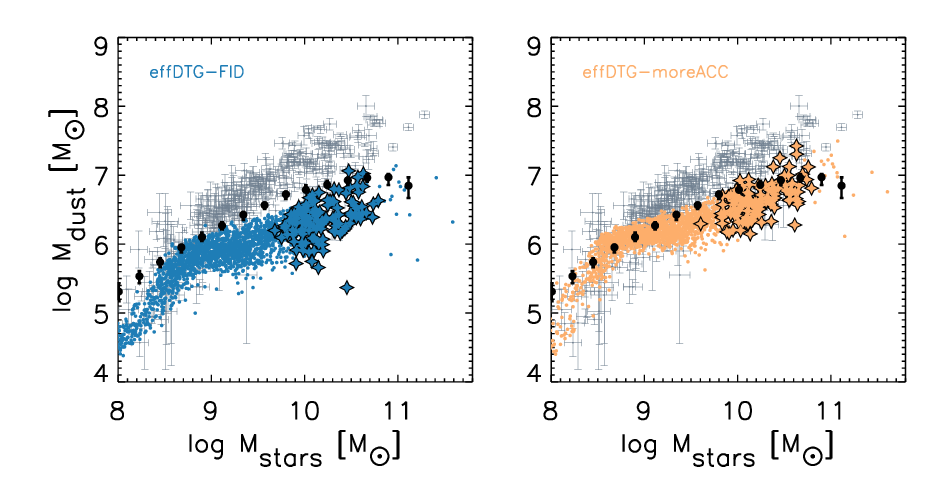}
    \caption{Stellar and dust mass at $z=0.0$ for the \textit{effDTG-FID} (blue, left panel) and \textit{effDTG-moreACC} model (yellow, right panel). Data from \citet{Beeston2018} and \citet{Vis2019} are shown for comparison. MWHM galaxies are marked with star symbols.}
    \label{fig:effdsg-MsMd}
\end{figure*}

\begin{figure*}
    \centering
    \includegraphics[width=1.5\columnwidth]{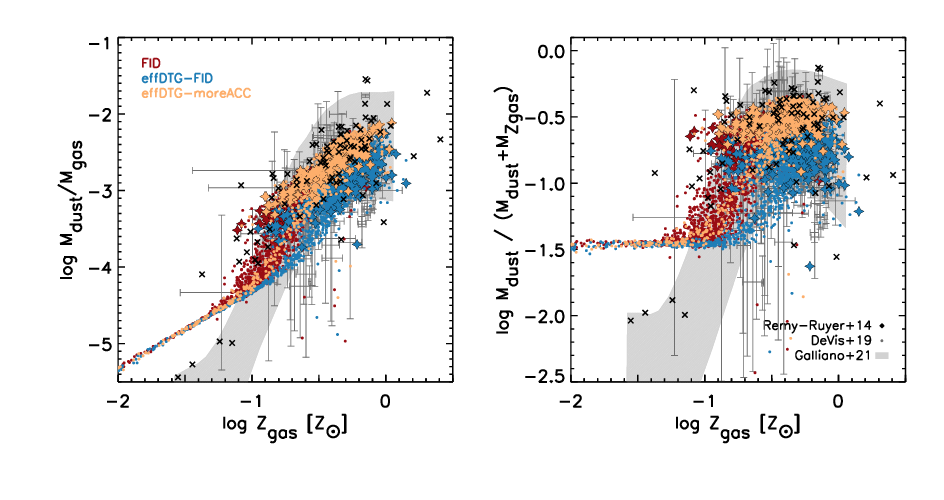}

    \caption{Dust-to-Gas (left panel) and Dust-to-Metals (right panel) ratio as a function of the ISM gas phase metallicity $Z$ for simulated galaxies at $z=0$ (MWHM galaxies are star symbols). The same three models as in Fig. \ref{fig:dustMF_effDTG} are shown (the fiducial of this work, and two with DTG dependent star formation efficiency). The compilations of observational data by \citet{Remy-Ruyer2014}, \citet{Vis2019}, and the fit by \citet{Galliano2021a} are shown for comparison.}
    \label{fig:effdsg-DustZ}
\end{figure*}


\begin{figure*}
    \centering
    \begin{subfigure}[t]{0.46\textwidth}
        \centering
        \includegraphics[width=\linewidth]{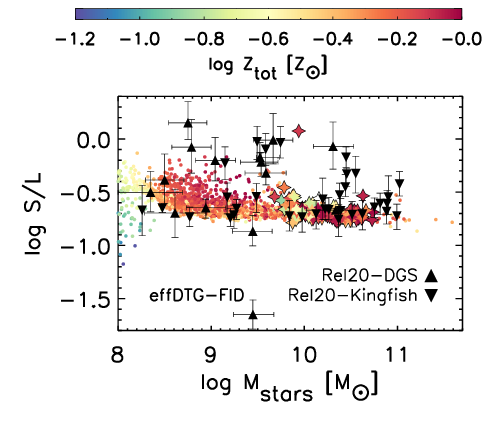}
    \end{subfigure}
    \hfill
    \begin{subfigure}[t]{0.46\textwidth}
        \centering
        \includegraphics[width=\linewidth]{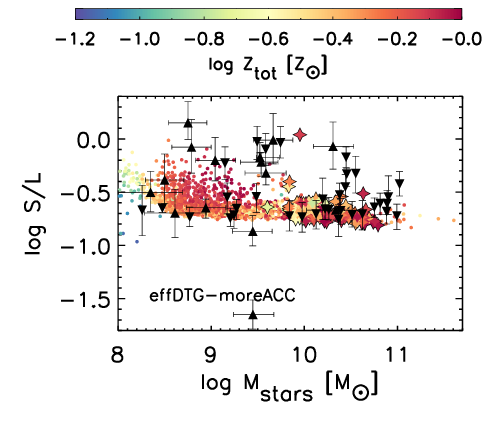}
    \end{subfigure}

    \caption{Small-to-Large grains ratio at $z=0.0$ as a function of the stellar mass for the \textit{effDTG-FID} (left) and \textit{effDTG-moreACC} (right) models. The total metallicity of each galaxy is color coded. Star symbols represent MWHM galaxies. Data from \citet{Relano2020} are shown for comparison (upward triangles for the DSG sample, downward triangles for Kingfish galaxies).}
    \label{fig:StoL_effDTG}
\end{figure*}

\begin{figure}
    \centering
    \includegraphics[width=1.\columnwidth]{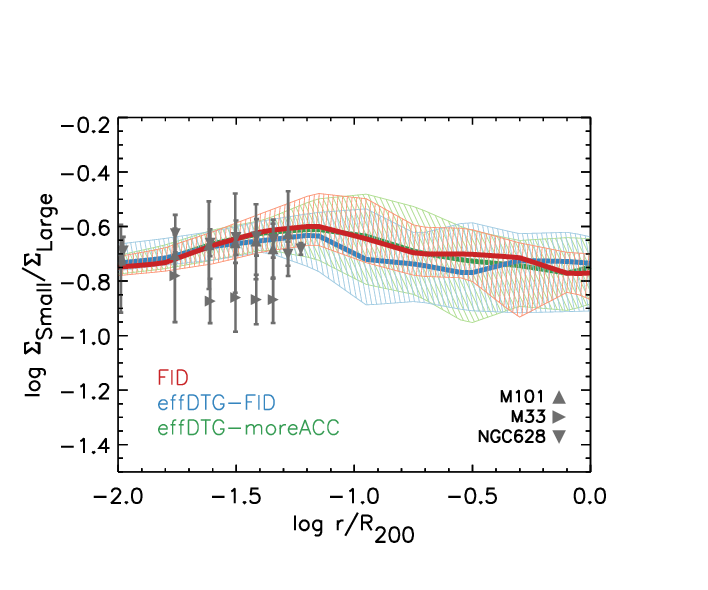}
    \caption{Median radial profiles ratio of Small-to-Large ratio for MWHM galaxies at $z=0.0$ for the fiducial (red), the \textit{effDTG-FID} (blue) and \textit{effDTG-moreACC} (green) model. The radial coordinate has been normalized to the $R_{200}$ of each galaxy. We show the 25-75th percentile dispersions. Observations by \citet{Relano2020} of three local galaxies (M101, M33, NGC628) are shown for comparison.}
    \label{fig:profileSL_effDTG}
\end{figure}

\begin{figure}
    \centering
    \includegraphics[width=1.\columnwidth]{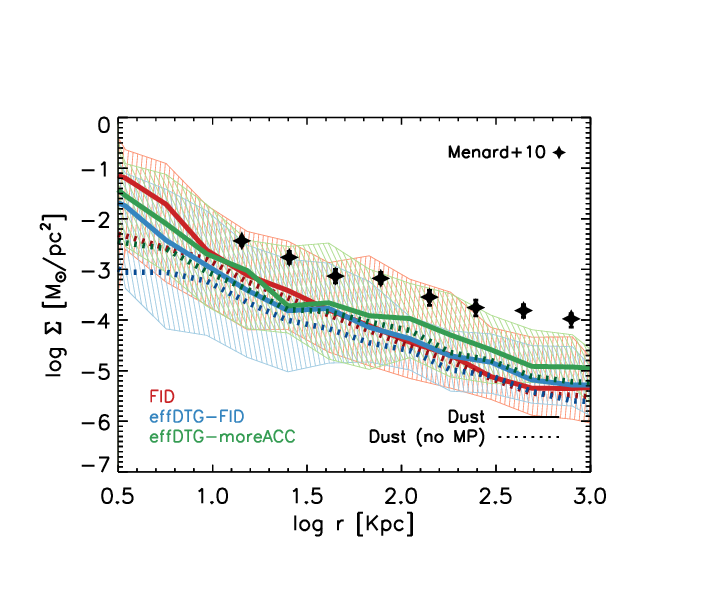}
    \caption{Median dust surface density profile (solid line) around galaxies with $M_\text{stars}>10^{10} \, M_\odot$ at $z=0.3$ for the fiducial (red), the \textit{effDTG-FID} (blue) and \textit{effDTG-moreACC} (green) model. The shaded area shows the full dispersion of the profiles distribution. The profile excluding MultiPhase gas particles is also shown as a dotted line. Filled star points data refer to observations by \citet{Menard2010}.}
    \label{fig:menard_effDTG}
\end{figure}

\label{app:effdsg}

\rev{We show some results obtained when adopting the DTG-dependent star formation efficiency discussed in Sec. \ref{sec:SFRD}, where we suggested as a viable possibility to improve the match with the observed cosmic star formation density.
We consider here two renditions of it. The run \textit{effDTG-FID} has dust parameters identical to the fiducial run, while in the \textit{effDTG-moreACC} the accretion efficiency has been enhanced by a factor $2.5$. The reasons for considering also the latter one will be clarified below.
The prediction for the cosmic \revdue{SFRD} was already included in Fig. \ref{fig:SFRD} (for \textit{effDTG-FID}, but \textit{effDTG-moreACC} is indistinguishable in this plot). The galaxy \revdue{MS} and the \revdue{SMF} (Fig. \ref{fig:mainsequence} and \ref{fig:stellarMF_z}) are statistically almost identical to those of the fiducial run, but for a relatively modest increase of the SMF by a factor $\sim 2$ at $M_{\rm stars}< 10^{10} M_{\odot}$. Therefore, we do not show them here.
As for the main properties of MWHM galaxies, discussed for the fiducial model in Section \ref{sec:mainpropMWgal}, \textit{effDTG} produces a moderate increase of the B-to-T ratio, particularly for the most massive MWHM galaxies, a factor $\sim 3$ decrease of the gas fraction $f_\text{gas}$, and a factor $\sim 2$ increase of the gas metallicity. These variations can be appreciated in Figure \ref{fig:BT_fg_Z_effDTG}.
The $f_\text{gas}$ decrease and the $Z_\text{gas}$ increase appear to be a natural consequence of the more efficient star formation, causing more gas consumption, stellar feedback and metal production. }

\rev{We focus now on dust-related predictions of the model.
Figs. \ref{fig:dustMF_effDTG},  \ref{fig:effdsg-MsMd} and \ref{fig:effdsg-DustZ} show the \revdue{DMF}, the $M_{\rm stars}-M_{\rm dust}$ and the DTG(DTM)-Z correlations. The results can be understood considering that, as remarked above, these modified models produce galaxies featuring a factor $\sim 3$ lower gas fraction than the fiducial one. An even higher decrease, by a factor $\sim 4$, occurs in the molecular gas fraction. As a result, the predicted molecular mass in MWHM galaxies in the \textit{effDTG} runs turns out to be less than $\sim 1.5$ \% of the baryonic mass, which is too low. Indeed, nearby star-forming late-type galaxies feature a typical fraction of $\gtrsim 3$ \% \cite[e.g.][]{Saintonge2022}. This ISM component is the dense gas in which accretion is assumed to occur \cite[see section 2.3.3 in][]{Granato2021}. Consequently, in the \textit{effDTG-FID} model, the accretion process is less productive than in the fiducial one, and the \textit{critical} metallicity (that above which accretion becomes relevant) shifts to larger values. Moreover, the dust content for a given stellar mass is lower and in clear tension with observations due both to the lower gas content and the lower \revdue{DTG} ratio. The most straightforward modification to recover the critical metallicity value (log $Z/Z_\odot \simeq -0.8$) suggested by observations and an acceptable dust to stellar mass ratio and \revdue{DMF} {is to increase the} accretion efficiency, as in \textit{effDTG-moreACC}. This adopted value is still twice as small as that we used by \cite{Granato2021}. Considering this and the fact that the \textit{effDTG} models require some future adjustment to increase the predicted molecular mass, the conclusion that the accretion timescale adopted in the latter work needs a downward revision appears to be robust.}

\rev{In both the models shown here, the transition region of the DTG-Z and DTM-Z diagrams between the stellar production and saturation regime is scarcely populated (see Fig. \ref{fig:effdsg-DustZ}). This feature arises from a faster evolution of galaxies. Due to the assumed relation between SFR and DTG, galaxies consume and reprocess gas faster when their DTG begins to increase significantly. The net result is that at $z=0$ many more objects than in the fiducial run are already in the saturation regime.}

\rev{The size and chemical composition of grains is not severely affected by the changes introduced by
\textit{effDTG} models. In particular, S/L as a function of $M_\text{stars}$ or $Z_\text{tot}$ and its radial profile are again in good keeping with observations (Figure \ref{fig:StoL_effDTG} and \ref{fig:profileSL_effDTG}).}

\rev{Finally, we show in Figure \ref{fig:menard_effDTG} the predictions for the dust profiles outside galaxies, discussed in Section \ref{sec:dustoutgalaxies}. While in the galactic region inside $\sim 30$ kpc the profiles predicted by the \textit{effDTG-FID} model lay below the fiducial one, consistently with the lower dust production in the former model, the differences tend to vanish at larger radii. The reason is that the increased SF activity and related feedback in the former model are more efficient in ejecting reprocessed gas, compensating for its lower dust production.}

\rev{The results presented in this appendix confirm that, although promising and relatively physically motivated, the DTG star formation efficiency suggestion requires a dedicated investigation.}

\bsp	
\label{lastpage}
\end{document}